\documentclass[conference]{IEEEtran}
\IEEEoverridecommandlockouts
\usepackage{cite}
\usepackage{amsmath,amssymb,amsfonts}
\usepackage{graphicx}
\usepackage{textcomp}
\usepackage{amsmath}
\usepackage{booktabs} 
\usepackage{cite}
\usepackage{multirow}
\usepackage{hyperref}  
\usepackage{stfloats}
\usepackage{soul}
\usepackage[font=small]{caption}

\usepackage{subcaption}
\usepackage{url}            
\usepackage{booktabs}       
\usepackage{amsfonts}       
\usepackage{nicefrac}       
\usepackage{microtype}      
\usepackage{xcolor}         
\usepackage{algorithm}
\usepackage{algpseudocode}
\usepackage{amsmath}
\usepackage{tikz}

\usepackage{orcidlink}
\usepackage{comment}
\usepackage{amssymb}
\usepackage{braket}
\usepackage{xcolor}
\def\BibTeX{{\rm B\kern-.05em{\sc i\kern-.025em b}\kern-.08em
    T\kern-.1667em\lower.7ex\hbox{E}\kern-.125emX}}
\begin{document}

\title{
Comparative Performance Analysis of Quantum Machine Learning Architectures for Credit Card Fraud Detection}


\author{\IEEEauthorblockN{Mansour El Alami\textsuperscript{1}, Nouhaila Innan\orcidlink{0000-0002-1014-3457}\textsuperscript{2,3}, Muhammad Shafique\orcidlink{0000-0002-2607-8135}\textsuperscript{2,3}, and Mohamed Bennai\orcidlink{0000-0002-7364-5171}\textsuperscript{1}
}
\IEEEauthorblockA{\textsuperscript{1}Quantum Physics and Spintronics Team, LPMC, Faculty of Sciences Ben M'sick,\\ Hassan II University of Casablanca,
Morocco\\
\textsuperscript{2}eBRAIN Lab, Division of Engineering, New York University Abu Dhabi (NYUAD), Abu Dhabi, UAE\\
\textsuperscript{3}Center for Quantum and Topological Systems (CQTS), NYUAD Research Institute, NYUAD, Abu Dhabi, UAE\\
mansour.elalami-etu@etu.univh2c.ma,  nouhaila.innan@nyu.edu, muhammad.shafique@nyu.edu,  mohamed.bennai@univh2c.ma \\
}}

\maketitle

\begin{abstract}
 As financial fraud becomes increasingly complex, effective detection methods are essential. Quantum Machine Learning (QML) introduces certain capabilities that may enhance both accuracy and efficiency in this area. This study examines how different quantum feature maps and ansatz configurations affect the performance of three QML-based classifiers, the Variational Quantum Classifier (VQC), the Sampler Quantum Neural Network (SQNN), and the Estimator Quantum Neural Network (EQNN), when applied to two non-normalized financial fraud datasets.
Different quantum feature map and ansatz configurations are evaluated, revealing distinct performance patterns. The VQC consistently demonstrates strong classification results, achieving an F1-score of 0.88, while the SQNN also delivers promising outcomes. In contrast, the EQNN struggles to produce robust results, emphasizing the challenges presented by non-standardized data. 
Statistical validation using ANOVA confirms the significance of observed performance differences. Additionally, robustness tests on the best-performing models under five quantum noise types show that they maintain competitive performance, supporting their practical applicability.
These findings highlight the importance of careful model configuration in QML-based financial fraud detection. By showing how specific feature maps and ansatz choices influence predictive success, this work guides researchers and practitioners in refining QML approaches for complex financial applications.
\end{abstract}

\begin{IEEEkeywords}
Quantum Machine Learning, Quantum Neural Networks, Fraud Detection, Quantum Variational Classifier
\end{IEEEkeywords}

\section{Introduction}
Credit card fraud represents a pervasive and escalating global threat, costing individuals and organizations billions of dollars each year. For instance, a 2022 UK Finance study reported fraud-related losses of £1.2 billion, with online solicitations accounting for approximately 80\% of fraudulent activities \cite{robinson2024fraudsters}. Similarly, a CNBC report published in 2023 estimated that fraud imposed an \$8.8 billion burden on American consumers in 2022 \cite{byrne2023scams}, while a joint assessment by the European Banking Authority and the European Central Bank indicated that credit card fraud reached 633 million euros in the first half of 2023 \cite{fraud2024}. These escalating figures threaten consumers and financial institutions, contributing to broader economic instability.
Machine Learning (ML) has been widely applied across domains for classification and decision tasks \cite{zhu2023multichannel, yang2024object}, including fraud detection. 
Conventional fraud detection methods have advanced significantly, yet they face persistent challenges. The increasing volume and complexity of transactional data, the highly imbalanced nature of fraud detection datasets, and the growing need for real-time decision-making continue to strain ML approaches \cite{baabdullah2024efficiency, mienye2024deep,abbassi2024real,thanathamathee2024shap}. Quantum Machine Learning (QML) offers a promising avenue to overcome these limitations \cite{schuld2015introduction}, potentially enhancing computational speed, handling complex data distributions more effectively, and improving overall detection accuracy. 
Moreover, QML has demonstrated success across a wide range of applications, extending from healthcare and finance to climate modeling \cite{dutta2024qadqn,innan2024lep,innan2025qnn}, materials science \cite{chen2024crossing,innan2024quantum}, and cybersecurity, illustrating its versatility in tackling complex \cite{maouaki2024quantum}, data-intensive problems \cite{pathak2024resource,innan2024quantum1}. This success largely depends on selecting and optimizing various QML models and architectural configurations designed to employ the unique capabilities of quantum computing \cite{innan2023enhancing,innan2024variational,10651123}. Consequently, developing effective QML-based solutions necessitates a thorough investigation of different model architectures and configurations to identify those that most effectively address the specific challenges of each application domain. In the context of financial fraud detection, this involves analyzing how different QML architectures impact performance metrics such as accuracy, recall, and F1-score, thereby ensuring the development of robust and efficient detection systems \cite{innan2024financial}.

This study focuses on examining how various QML configurations, specifically, configurations of feature maps and ansatz architectures, influence accuracy, recall, and F1-score when applied to two real-world credit card fraud detection datasets. In addition, we incorporate statistical validation of performance through Analysis of Variance (ANOVA) and evaluate the robustness of the most effective models under different quantum noise scenarios. By evaluating a range of model setups and pinpointing those configurations that achieve the best performance, this work provides valuable insights for designing more effective QML-based fraud detection systems. Our study further provides a systematic multi-factor analysis of QML architectures, exploring how model type, feature map, ansatz design, and dataset characteristics interact and affect performance. This broader evaluation uncovers dependencies that are often overlooked in prior work and clarifies how these architectural components collectively influence fraud detection outcomes. Ultimately, this analysis contributes to the broader goal of protecting consumers and businesses from increasingly sophisticated financial threats.
        
\begin{figure*}[t]
    \centering
    \includegraphics[width=\linewidth]{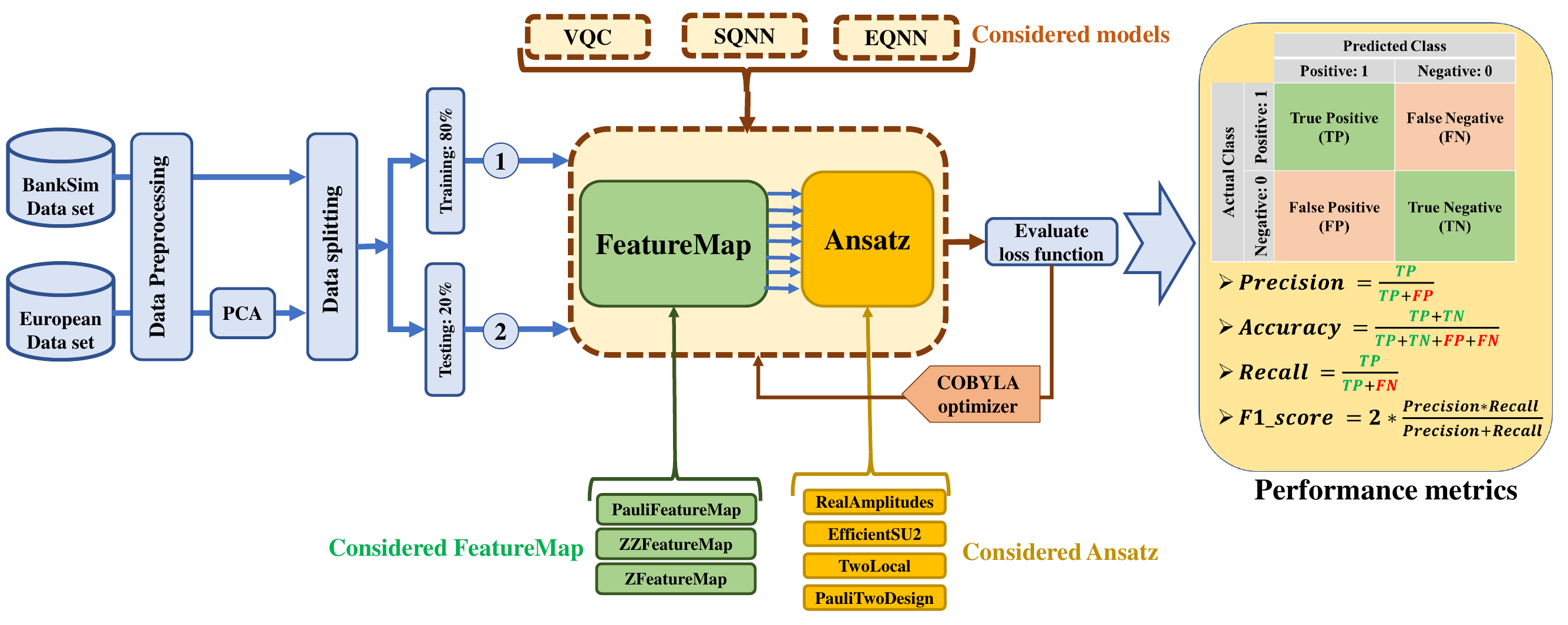}
    \caption{Overview of the proposed methodology. Two benchmark datasets undergo initial preprocessing, after which three QML models are evaluated. Each model is tested with three distinct feature map architectures, and each selected feature map is further assessed using four different ansatz configurations.}
    \label{Experimental Setup}
\end{figure*}

\textbf{The contributions of this work can be summarized as follows (as illustrated by the overall methodology in Fig. \ref{Experimental Setup}):}
\begin{itemize}
    \item A comprehensive analysis of three emerging QML models, Variational Quantum Classifier (VQC), Sampler Quantum Neural Network (SQNN), and Estimator Quantum Neural Network (EQNN), applied to credit card fraud detection.
    \item A thorough empirical evaluation of two real-world credit card fraud datasets, examining how each QML model behaves across key performance metrics.
   \item An in-depth examination of how feature maps and ansatz configurations influence model accuracy and detection capability, showing how encoding choices and circuit expressivity affect performance.
\item A rigorous statistical validation using ANOVA to measure the individual and interaction effects of model type, dataset, feature map, and ansatz on F1-score, clarifying which design choices drive performance.
    \item Insights into optimal feature map–ansatz combinations that deliver the strongest results for quantum-enhanced fraud detection.
    \item A robustness analysis of the best-performing model from each dataset, evaluated under five types of quantum noise to assess resilience and transferability to real NISQ hardware conditions.
\end{itemize}

Through these contributions, we explain the critical parameters that influence the performance of these QML models and the impact of non-standard financial fraud detection datasets.

The rest of the paper is structured as follows. In Sec.~\ref{sec2}, we review the existing literature on classical and QML approaches to financial fraud detection. In Sec.~\ref{sec3}, we provide the background and the architecture of different feature map and ansatz architectures. In Sec.~\ref{32}, we describe the QML models employed and the associated circuit constructs. In Sec.~\ref{sec4}, we present the datasets used, report our experimental results, and highlight the model configurations achieving the highest performance. Finally, in Sec.~\ref{sec5}, we conclude by summarizing our findings and discussing potential directions for future work.

\section{Literature Review\label{sec2}}

 Classical ML algorithms have been extensively explored for detecting fraudulent credit card transactions. A variety of models, including multivariate logistic regression \cite{alenzi2020fraud}, support vector machines \cite{kumar2022credit}, and random forest classifiers \cite{xuan2018refined,xuan2018random}, have demonstrated varying degrees of success. More sophisticated approaches integrate intelligent techniques, such as optimized light gradient boosting machine \cite{taha2020intelligent}, as well as deep learning architectures like auto-encoders and restricted Boltzmann machines \cite{pumsirirat2018credit}, and graphical neural networks \cite{ma2021comprehensive}.
 
Despite significant progress in classical approaches, these methods face persistent challenges due to the volume, complexity, and continuously evolving patterns of fraudulent activities. This has motivated interest in QML, which employs quantum properties—including entanglement and superposition—to operate efficiently in high-dimensional feature spaces. As a result, QML has the potential to process large datasets more rapidly, capture complex interactions among transactional features, and express intricate patterns, offering tangible benefits for fraud detection tasks \cite{peral2024systematic, tudisco2024evaluating}. 

The emergence of accessible QML tools, such as IBM's Qiskit and Xanadu's PennyLane \cite{qiskit, pennylane}, has further accelerated research in this area.
Several studies have begun to explore QML solutions for fraud detection. For instance, Liang \textit{et al. } introduced two quantum anomaly detection algorithms grounded in density estimation and multivariate Gaussian distributions that can be applied to fraudulent transaction identification \cite{liang2019quantum}. A quantum graph convolutional neural network proposed expanded quantum neural network to graph-structured data \cite{zheng2021quantum}, holding promise for fraud scenarios involving complex relational information. Another study compared quantum kernel-based anomaly detection with classical baselines such as one-class support vector machines, demonstrating the potential of QML in improving detection performance \cite{kyriienko2022unsupervised}.

Additional research has directly compared quantum-enhanced methods to classical benchmarks. Grossi \textit{et al.} evaluated a quantum support vector machine against traditional models like XGBoost and random forests within a financial fraud detection setting \cite{grossi2022mixed}, while Shetakis \textit{et al.} assessed quantum classifiers alongside classical artificial neural networks in a hybrid framework \cite{schetakis2024quantum}. Explorations into more specialized QML architectures include investigations of quantum neural networks for graph-structured quantum data in Noisy Intermediate-Scale Quantum (NISQ) devices \cite{beer2023quantum}. Recently, Quantum Graph Neural Networks (QGNN) have been employed to detect financial fraud, showcasing improvements over their classical GNN counterparts \cite{innan2024qgnn}.

Beyond algorithmic innovations, hybrid frameworks have emerged; Innan \textit{et al.} integrated QML with federated learning to propose quantum federated artificial neural networks for financial fraud detection \cite{innan2024qfnn}, improving both privacy and detection effectiveness. Furthermore, Huot \textit{et al.} introduced a QML model based on quantum auto-encoders for fraud detection \cite{huot2024quantum}, illustrating the adaptability of QML to various architectural paradigms.
In summary, the literature demonstrates the potential of QML in enhancing credit card fraud detection. Quantum methods improve detection rates and reduce false positives, offering more robust protection for consumers and financial institutions. As research progresses, refining and optimizing QML architectures will be crucial, ensuring these emergent techniques can reliably outperform their classical counterparts and contribute to safer financial ecosystems.

\section{Background\label{sec3}} 
\begin{figure}[htpb]
			\centering
         \includegraphics[width=1\linewidth]{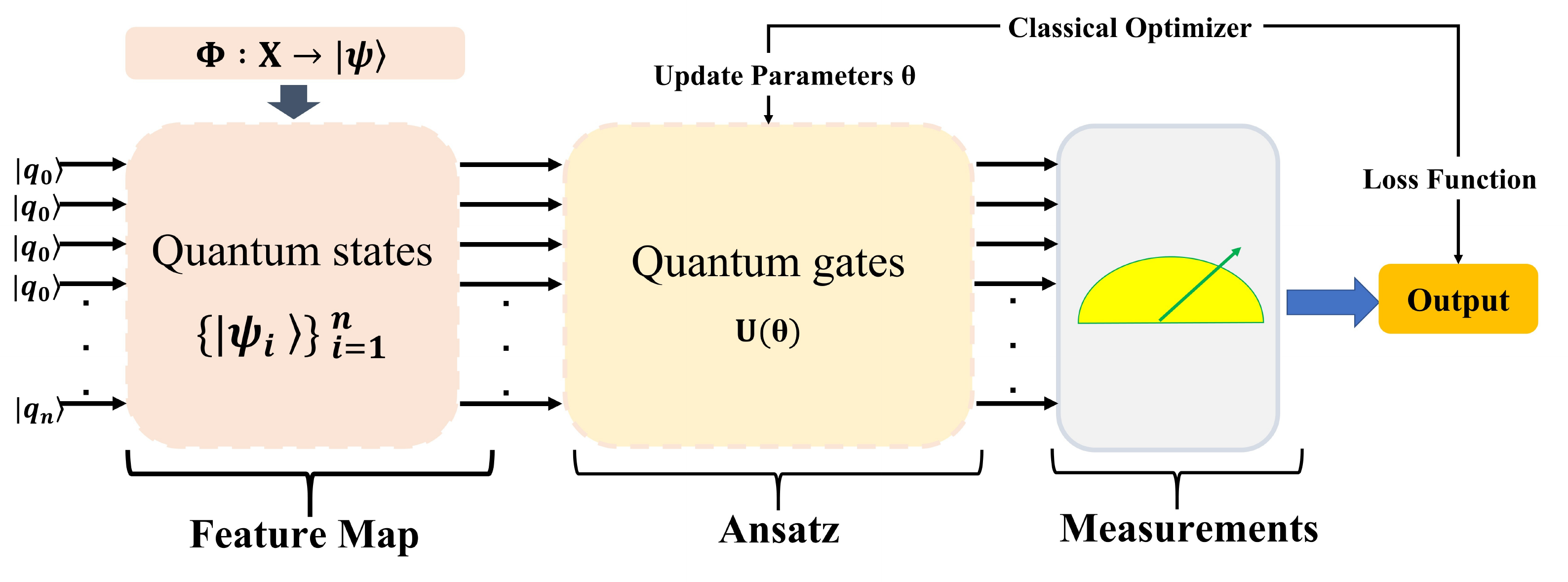}
			\caption{Fundamental architecture of a QML model. It involves transforming classical data into quantum states $|\psi_i\rangle$, processing them through an ansatz $\mathbf{U}(\theta)$, and optimizing its parameters using a classical optimizer based on a loss function until convergence is achieved iteratively. The final quantum state is measured to generate predictions.}
            
            \label{fig:QML-models.pdf}
		\end{figure}

\begin{figure}[!h]
			\centering
			\includegraphics[width=\linewidth]{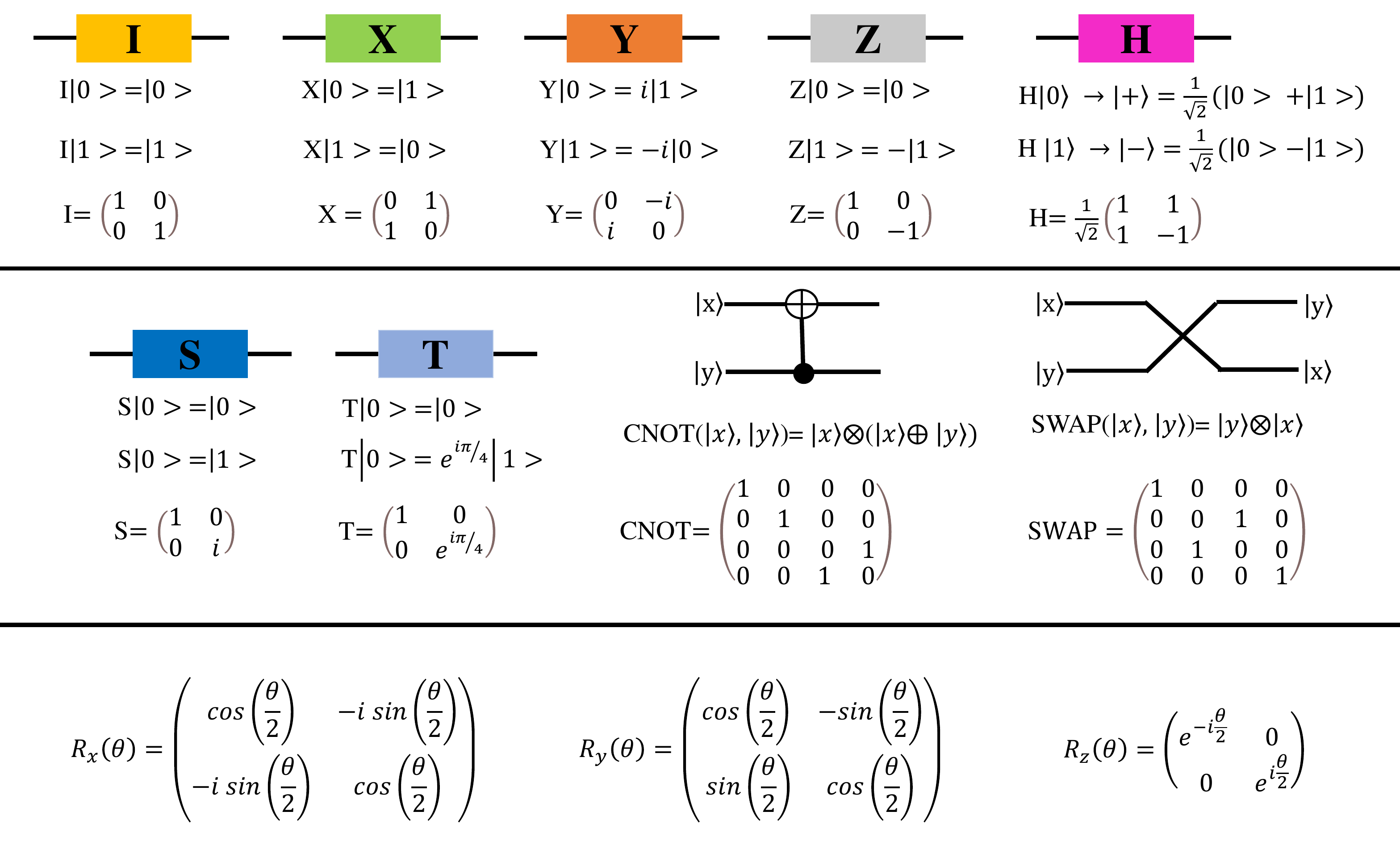}
            \vspace{-0.5cm}
			\caption{Graphical representation and mathematical expressions of selected single and two-qubit quantum gates.}
            \label{Quantum gates.pdf}
		  \end{figure}

The functionality of a QML model, as illustrated in Fig.~\ref{fig:QML-models.pdf}, is based on an architecture that comprises three fundamental components. Initially, the feature map encodes the input features of the dataset, denoted as ${|\psi_i\rangle}_{i=1}^n$, using an encoding scheme $\Phi: x \rightarrow |\psi\rangle$.
Subsequently, the ansatz $\mathbf{U}(\theta)$ applies trainable parameters $\theta$ through a sequence of quantum gates, evolving the quantum states of the system. 
The final component involves measurement and optimization, where the evolved quantum states are measured to extract quantum information. This information is then processed by a classical optimization algorithm, which iteratively adjusts the parameters $\theta$ to minimize the loss function, thereby enhancing the QML model's performance. This optimization loop effectively bridges the quantum and classical domains, ensuring continuous refinement of the model's predictive accuracy.
Additionally, Fig.~\ref{Quantum gates.pdf} shows the quantum gates commonly employed in the quantum circuits of QML models, highlighting their role in manipulating quantum states to perform the necessary computations for learning and optimization tasks. 

\subsection{Feature Maps}

A quantum feature map encodes the classical input vector $\vec{x}$ into a quantum state $|\psi(\vec{x})\rangle$ in a Hilbert space $F$ through the mapping $x \mapsto |\psi(\vec{x})\rangle$. This feature map, also known as the unitary operator $U_\psi(\vec{x})$, generates the quantum state as ${|\psi(\vec{x})\rangle = U_\psi(\vec{x}) |0\rangle}$. It acts on the ground or vacuum state $|0\rangle = |00\ldots0\rangle$ of the Hilbert space $F$. Using this approach, the feature space becomes more complex, enabling a quantum classification system to identify intricate patterns in the input data \cite{schuld2019quantum}. Various types of feature maps exist; we focus on angle encoding, specifically the Pauli, ZZ, and Z feature maps.

\subsubsection{Pauli Feature Map}
The Pauli feature map represents a quantum circuit that enables a Pauli expansion of a given dataset. The Pauli expansion is a method for representing the dataset as a product of Pauli operators, where each Pauli operator corresponds to a distinct feature within the data. The expression for the combination of Pauli operators is described as follows:

\begin{equation}
U_{\varphi_s(\vec{x})} = \exp \left( i \sum_{S \subseteq [n]} \varphi_s(\vec{x}) \prod_{i \in S} P_i \right),
\label{eq:2}
\end{equation}
where \(i\) is the set of qubit indices describing the connections in the feature map, \(P_i \in \{I, X, Y, Z\}\) denotes the Pauli matrices, and $S$ describes the correlation between different qubits. The set \( S \) belongs to the set of all combinations \( S \in \left\{ \binom{n}{k}, \; k = 1, \dots, n \right\} \). The term \(\varphi_s(\vec{x})\) represents the data mapping function \cite{alomari2023deqsvc}.

The data mapping function, \(\varphi_s(\vec{x})\), maps classical input data \(\vec{x}\) into the quantum circuit, enhancing the circuit's representational capabilities. It is defined as follows:
\begin{equation}
\varphi_s(x) =
\begin{cases} 
x_0 & \text{si } s = \{i\} \\
\prod_{j \in S} (\pi - x_j) & \text{si } |S| > 1 
\end{cases}.
\label{eq:3}
\end{equation}
The Pauli feature map circuit is constructed by initially applying Hadamard gates to all qubits. Subsequently, a series of rotation gates are applied to the qubits, with the rotation angle of each qubit determined by the data mapping function \(\varphi_s(\vec{x})\). Finally, entangling gates are applied to the qubits. 

\subsubsection{ZZ Feature Map}
The ZZ feature map is a specific implementation of the Pauli-Z operator. Representing a second-order Pauli-Z evolution, the ZZ operator is defined as: 
\begin{equation}
ZZ = \begin{pmatrix} 1 & 0 \\ 0 & -1 \end{pmatrix} \otimes \begin{pmatrix} 1 & 0 \\ 0 & -1 \end{pmatrix} = 
\begin{pmatrix}
1 & 0 & 0 & 0 \\
0 & -1 & 0 & 0 \\
0 & 0 & -1 & 0 \\
0 & 0 & 0 & 1
\end{pmatrix},
\label{eq:zz}
\end{equation}
which captures pairwise correlations between qubits. The circuit begins with Hadamard gates applied to all qubits, followed by a sequence of rotation and entanglement blocks. The rotation block utilizes a classical non-linear function, $\varphi(x)$, to determine rotation angles based on the input data; in this case, for example, seven features can be described as follows:
\begin{equation}
\varphi(x_0, x_1, \ldots, x_6) = \prod_{i=0}^{6} (\pi - x_i).
\end{equation}

Entanglement blocks are constructed using CNOT gates that entangle qubits according to a predefined structure. Repeated layers of rotation and entanglement blocks allow the circuit to encode increasingly complex correlations. The feature dimension corresponds to the number of qubits, and the repetition criterion determines the number of layers applied, making the ZZ feature map adaptable to the input data's dimensionality and the desired circuit depth.

\subsubsection{Z Feature Map}
The Z feature map class represents a first-order Pauli-Z evolution circuit. As it is a type of Pauli feature map and operates with fixed ``Z'' Pauli chains, its first-order expansion does not contain entangled gates. This feature map is especially well-suited to certain applications where a shallow, entanglement-free quantum circuit is required because of this special feature. Similar to the ZZ feature map, the Z feature map is customized to fit a specific number of qubits, known as the feature dimension, and allows the user to choose how many times the rotation blocks should be replicated. The circuit is built by applying Hadamard gates to each qubit, then rotation blocks. The rotation blocks are structured using the same methods used in the ZZ feature map.

The Z feature map encoding method also offers parameters like repetition count, and entanglement approach, crucial for circuit inspection. Since there are no entanglement gates in the Z feature map, the entanglement strategy is zero. By offering a different quantum feature map that fits particular usage scenarios where entangled gates must be avoided, this feature map enhances the ZZ feature map method.

\subsection{Ansatz}
An ansatz refers to a parameterized quantum circuit designed to approximate a solution to a specific problem or encode a given dataset into quantum states. It consists of a sequence of quantum gates with tunable parameters, enabling the circuit to explore various quantum states. By adjusting these parameters, the ansatz can be optimized to minimize a cost function or represent features of a given problem. The design of an ansatz is crucial in QML, as it determines the expressiveness and trainability of the model. Tailoring the ansatz to the specific problem can incorporate problem-relevant symmetries and constraints, improving efficiency and solution quality \cite{cheng2022topgen}.

\begin{figure}[htpb]
    \centering
    \includegraphics[width=1\linewidth]{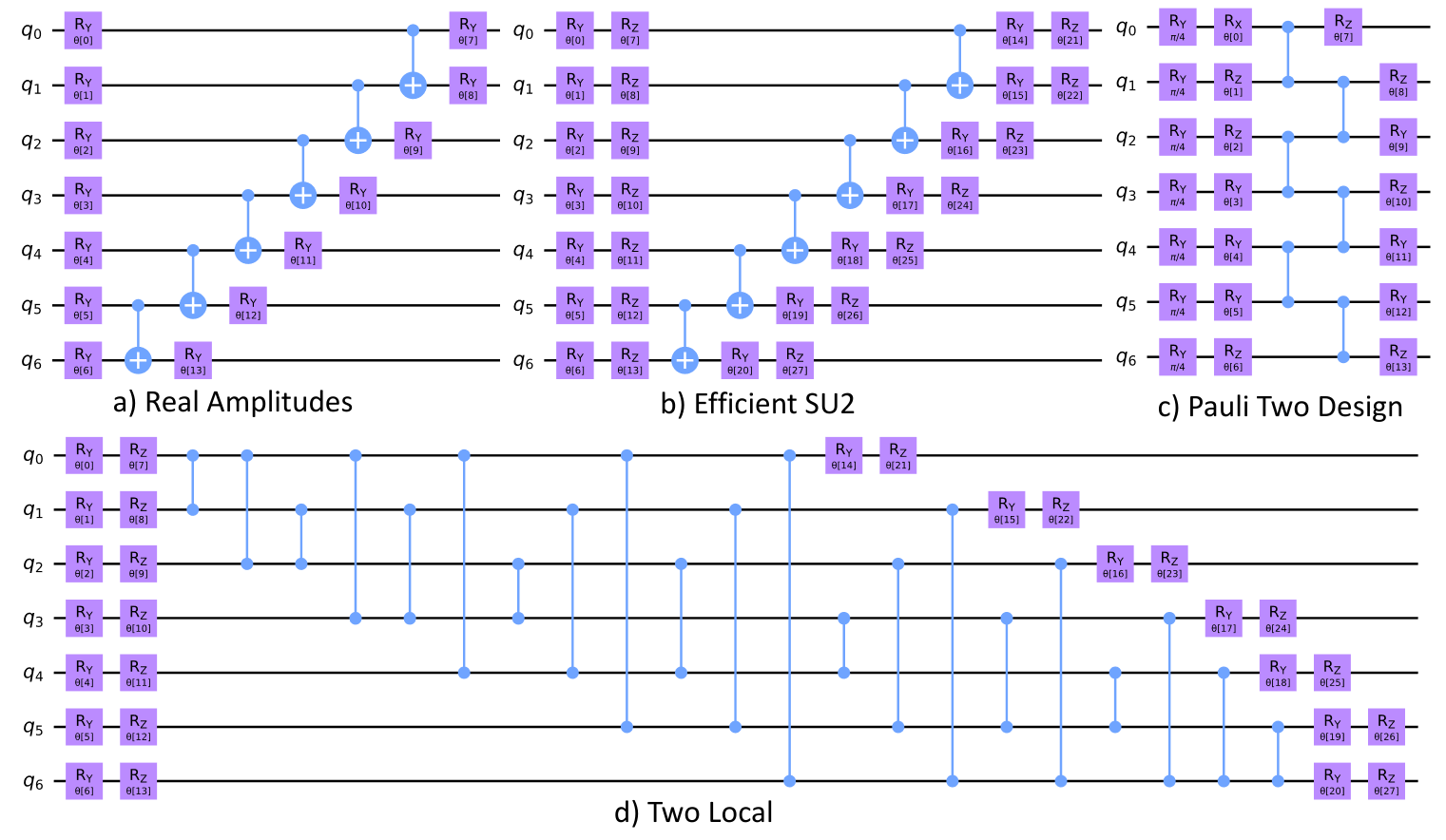}
    \caption{Architectures of the four ansatz variants utilized in this study, demonstrated on seven qubits: \textbf{a)} Real Amplitudes, \textbf{b)} Efficient SU2, \textbf{c)} Two Local, and \textbf{d)} Pauli Two Design, highlighting their structural differences and configurations.}
    \label{fig:ansatz}
\end{figure}

This study focuses on four distinct ansatz architectures: Real Amplitudes, Efficient SU2, Two Local, and Pauli Two Design. These architectures are chosen for their diverse properties and capabilities in capturing correlations and representing quantum states, as illustrated in Fig.~\ref{fig:ansatz}.

\subsubsection{Real Amplitudes}
The Real Amplitudes circuit serves as a heuristic test wave function. It comprises alternating layers of CNOT entanglement gates and \( Y \)-rotation gates, as illustrated in Fig.~\ref{fig:ansatz}-a. The entanglement pattern can be pre-selected from standard configurations or customized by the user. Notably, the quantum states generated by this circuit have exclusively real amplitudes, with zero complex components, giving the circuit its name.
  
\subsubsection{Efficient SU2}
The Efficient SU2 circuit consists of alternating layers of single-qubit \( \mathrm{SU}(2) \) gates and CNOT entanglement gates, as shown in Fig.~\ref{fig:ansatz}-b. The \( \mathrm{SU}(2) \) group encompasses \( 2 \times 2 \) unitary matrices with a determinant of 1, such as Pauli rotation gates.
\subsubsection{Pauli Two Design}
The Pauli Two Design circuit begins with an initial layer of gates, \( \sqrt{H} = \mathbf{R}_y\left(\frac{\pi}{4}\right) \), followed by alternating layers of single-qubit rotation gates and entanglement gates, as shown in Fig.~\ref{fig:ansatz}-c. In the rotation layers, single-qubit Pauli rotations are applied along axes \( X \), \( Y \), or \( Z \), selected uniformly at random. The entanglement layers consist of controlled-Z (CZ) gates arranged in a matching pattern.

\subsubsection{Two Local}
The Two Local circuit is a parametric quantum circuit composed of alternating rotation and entanglement layers. The rotation layers apply single-qubit gates to all qubits, while the entanglement layers employ two-qubit gates to establish entanglement according to a predefined strategy, as represented in Fig.~\ref{fig:ansatz}-d. The specific rotation and entanglement gates can be defined explicitly, such as \( R_Y \) and CNOT.

\section{QML Models\label{32}}
\subsection{Variational Quantum Classifiers}
\begin{figure}[!h]
			\centering
			\includegraphics[width=\linewidth]{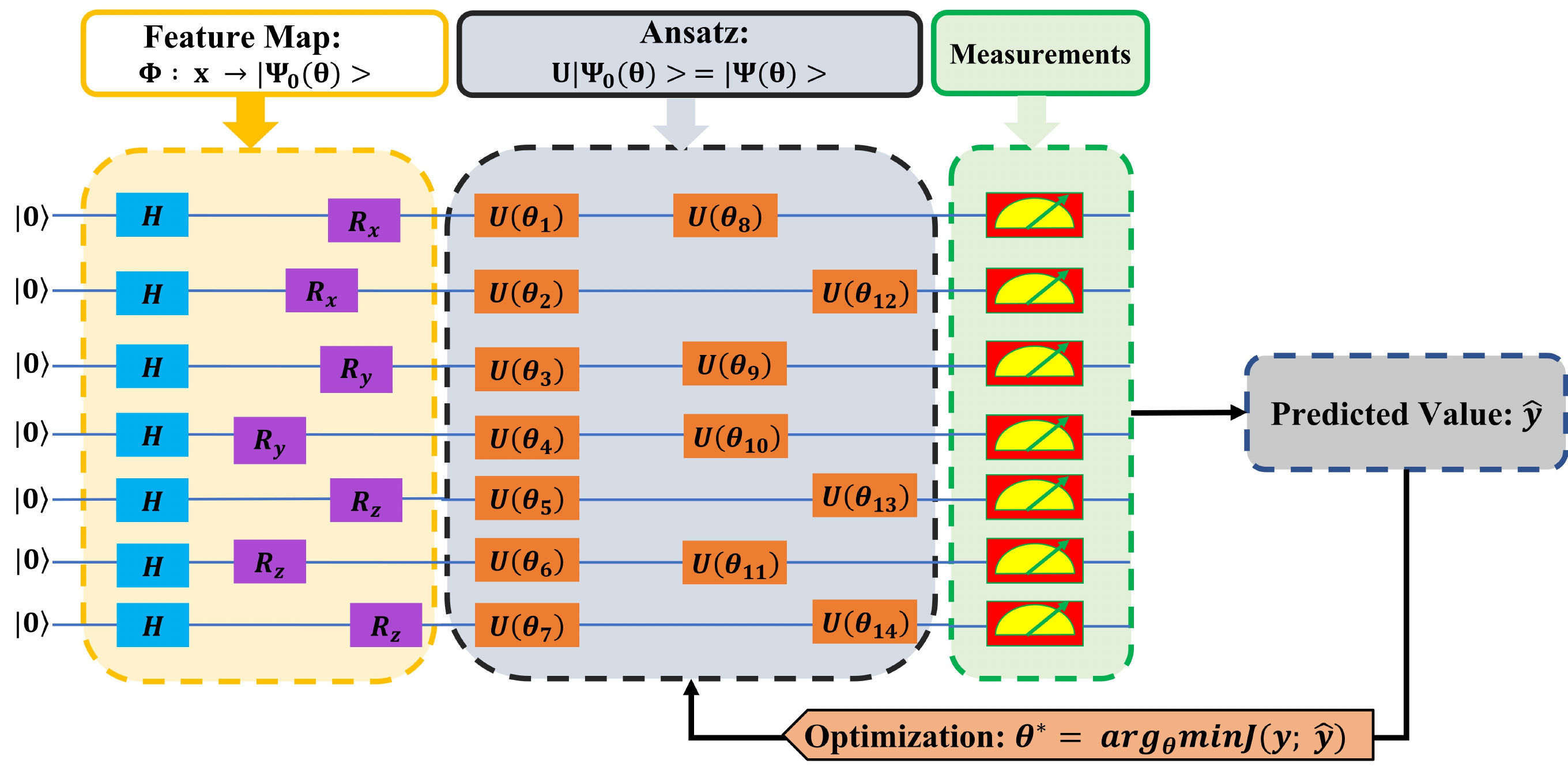}
			\caption{Architecture of the VQC model, it consists of a feature map, the $U$ operator is then applied in a measurement to obtain the classical output $\hat{y}$. A classical optimizer iteratively updatethe parameters $\theta$ to minimize the loss function $J(y; \hat{y})$.}
            \label{fig:VQC Model}
		  \end{figure}

The VQC is a parameterized quantum circuit whose parameters, referred to as learnable weights, are optimized using classical methods to minimize a loss function. The operation of a VQC, as shown in Fig.~\ref{fig:VQC Model}, proceeds as follows:
First, the quantum state preparation step initializes the circuit with parameters \(\boldsymbol{\theta} = \left(\theta_1, \theta_2, \dots, \theta_n\right)\), where \(n\) is the number of registers in the circuit. These parameters represent the learnable weights, which are initially set to random values in the range \(0 \leq \theta_i \leq 1\). The initial quantum state denoted as \(\left|\Psi_0(\theta)\right\rangle\), typically consists of simple quantum states, such as a tensor product of \(\left|0\right\rangle\) states.

Next, a unitary transformation is applied. This step involves the application of a sequence of quantum gates to the initial state. 
Let \(G_i \in \left\{I, X, Y, Z, H, S, T, \mathrm{R}_X, \mathrm{R}_Y, \mathrm{R}_Z, \mathrm{CNOT}, \mathrm{SWAP}, \dots\right\}\) represent the set of possible quantum gates, where \(1 \leq i \leq m\). The gates are applied to form a unitary operator \(U = \bigotimes_{i=1}^m G_i\). The resulting quantum state after the unitary transformation is given by:
\begin{equation}
U\left|\Psi_0(\theta)\right\rangle = \left|\Psi(\theta)\right\rangle.
\end{equation}

Finally, measurement is performed on the transformed quantum state \(\left|\Psi(\theta)\right\rangle\) to extract information and compute a classical output. Steps 2 and 3 are iteratively executed to minimize the loss function \(J(\theta)\). The process terminates when an acceptance criterion, such as convergence of the loss function, is met.

\subsection{Estimator Quantum Neural Networks}
\begin{figure}[htpb]
			\centering			
            \includegraphics[width=\linewidth]{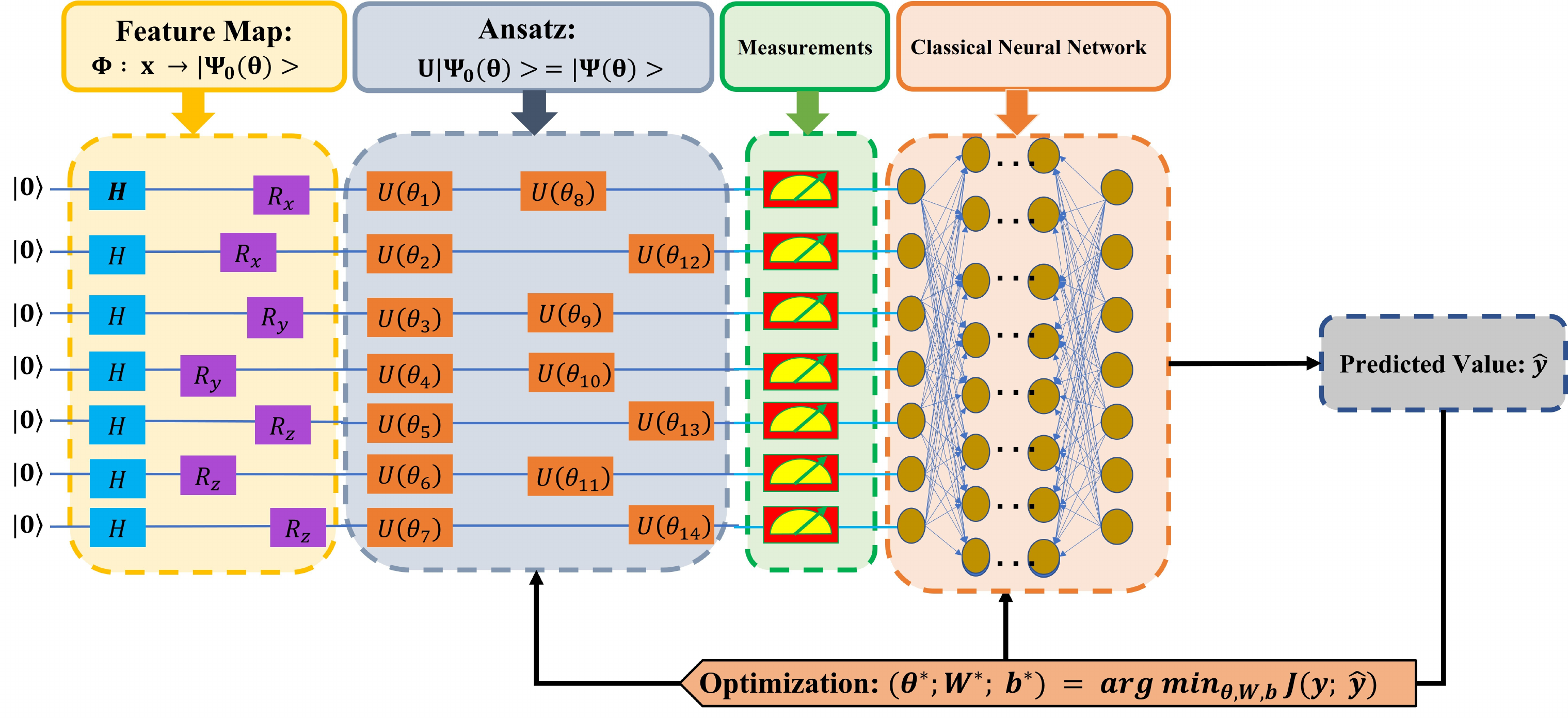}
			\caption{Architecture of the EQNN model. The process starts with a feature map, processing it via ansatz, then feeds the measured output into a classical neural network. A classical optimizer updates both quantum and neural network parameters ($\theta^*$, $\mathbf{W}^*$, and $b^*$) to minimize the loss function $J(y, \hat{y})$.}
            \label{fig:EQNN Model}
		\end{figure}

The EQNN is a hybrid architecture that integrates classical and QNNs. As shown in Fig. \ref{fig:EQNN Model}, the EQNN is composed of three primary components: a quantum feature map, an ansatz including the measurement, and a classical neural network.

The operation of the EQNN begins with data preparation via a quantum feature map. Classical data, denoted as $\mathbf{x} = (x_1, x_2, \dotsc, x_n)$, is encoded into quantum states $\left|\Psi_0(\theta)\right\rangle$ through the quantum feature map $\Phi: \mathbf{x} \mapsto \left|\Psi_0(\theta)\right\rangle$. Following the encoding stage, the quantum states $\left|\Phi(\mathbf{x})\right\rangle$ are processed through the ansatz to transform the encoded states into output quantum states $\left|\Psi(\theta)\right\rangle$. 
Subsequently, measurements are performed on the output states $\left|\Psi(\theta)\right\rangle$, extracting classical features from the quantum system. These features are obtained by measuring the qubits in the computational basis ${\left|0\right\rangle, \left|1\right\rangle}$.
The classical features obtained from the measurement stage are then processed by a fully connected classical neural network, which computes the output values $\hat{y}$. This step combines the strengths of classical and quantum processing for robust learning.
Finally, the model undergoes an optimization phase where the parameters of both the quantum and classical components are adjusted. The goal is to determine the optimal parameters $\theta^*$ for the ansatz and the weights $\mathbf{W}^*$ and biases $b^*$ of the classical neural network by minimizing the loss function $J(y; \hat{y})$. This process can be formulated as:
\begin{equation}
(\theta^*; \mathbf{W}^*; b^*) = \arg\min_{\theta, \mathbf{W}, b} J(y; \hat{y}).
\end{equation}
The optimization of quantum and classical parameters is conducted simultaneously, ensuring efficient learning and improved performance.

\subsection{Sampler Quantum Neural Networks}
\begin{figure}[!h]
			\centering
			\includegraphics[width=\linewidth]{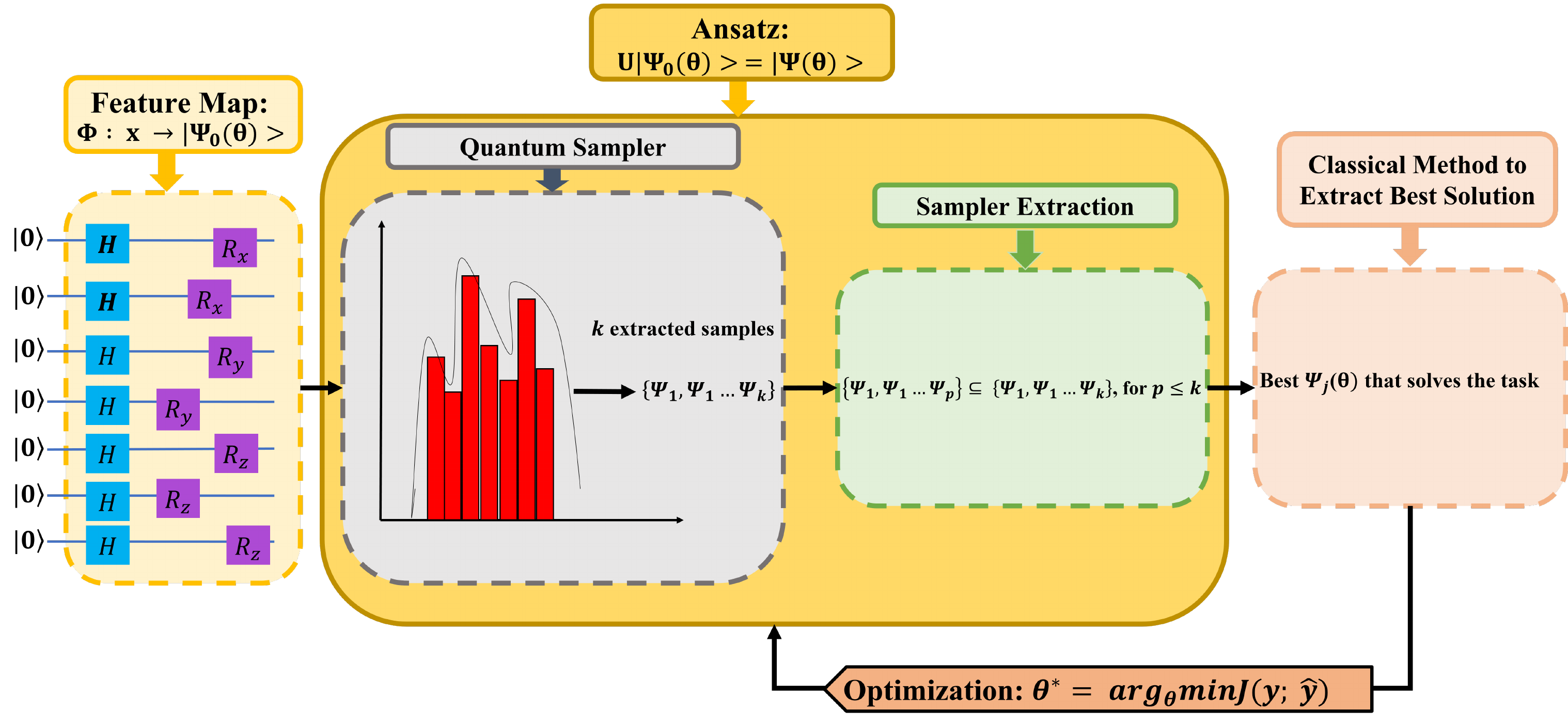}
			\caption{Architecture of the SQNN model. It consists of a feature map and an ansatz that incorporates a quantum sampler, followed by the extraction of samples generated by the quantum sampler.}
            \label{fig:SQNN Model}
		\end{figure}

The SQNN, similar to the EQNN, also features a hybrid classical-quantum architecture. However, the key distinction of the SQNN lies in the inclusion of a quantum sampler, which extracts examples of quantum states based on the complex probability distribution associated with these states, as shown in Fig. \ref{fig:SQNN Model}.

The SQNN operates through several stages. First, the process begins with the generation of quantum states via a feature map identical to the first step in the EQNN model. Next, a quantum sampler is applied. This step aims to efficiently extract examples of quantum states corresponding to the complex probability distribution of the problem's solution in a specific variable configuration. 

Following this, samples are extracted from the examples generated by the quantum sampler. The extracted samples are then processed using classical methods to identify the best solutions, where the optimal solutions for the given task are selected using a classical optimization scheme. Finally, a classical optimization method is employed to find the optimal parameters by minimizing the loss function, expressed as:
\begin{equation}
\theta^* = \arg\min_{\theta} J(y, \hat{y}),
\end{equation}
where the optimized $\theta$ values are returned to the quantum circuit. The entire process is repeated iteratively until the optimal parameter values are determined, ensuring the model's convergence to the best solutions for the given problem.

\section{Results and Discussion\label{sec4}}
\subsection{Dataset and Feature Selection}
The first dataset used in this study is derived from BankSim \cite{EalaxiKaggle2017}, an agent-based bank payment simulator based on aggregated transactional data provided by a leading bank in Spain. BankSim is designed to generate synthetic data tailored for fraud detection research. To achieve this, statistical analysis and social network analysis are utilized to study relationships between merchants and customers, resulting in a calibrated model \cite{LopezRojasAxelsson2014}. 

The BankSim dataset comprises 594,643 records obtained over 180 simulation steps, equivalent to approximately six months of activity. Of these records, 587,443 are regular payments, while 7,200 represent fraudulent transactions. The simulated frauds are introduced by modeling thieves stealing an average of three cards per stage and conducting around two fraudulent transactions per day. This dataset includes nine feature columns and one target column, which are essential for analyzing patterns and characteristics. The features are as follows:

\begin{itemize}
    \item \textbf{``Step''}: Represents the time aspect, designating the simulation day, spanning 180 steps over six months.
    \item \textbf{``Customer''}: Identifies individual customers involved in transactions.
    \item \textbf{``ZipCodeOrigin''}: Indicates the origin postcode of each transaction, facilitating geographical analysis.
    \item \textbf{``Merchant''}: Identifies merchants involved in transactions.
    \item \textbf{``ZipMerchant''}: Provides the postcode associated with each merchant.
    \item \textbf{``Age''}: Categorizes customer age into bands: '0' ($\leq 18$), '1' (19–25), '2' (26–35), '3' (36–45), '4' (46–55), '5' (56–65), '6' ($>65$), and 'U' (Unknown).
    \item \textbf{``Gender''}: Categorizes customers as ``E'' (Enterprise), ``F'' (Female), ``M'' (Male), or ``U'' (Unknown).
    \item \textbf{``Category''}: Captures the purchase category, indicating the type of transaction.
    \item \textbf{``Amount''}: Represents the monetary value of transactions.
    \item \textbf{``Fraud''}: The target variable, a binary class where 1 indicates fraud and 0 indicates non-fraud.
\end{itemize}

As part of preprocessing, inconsistencies in the ``Age'' column are resolved by converting categorical values into integers using regular expressions. Categorical features such as ``Customer'', ``Gender'', ``Merchant'', and ``Category'' are numerically encoded for efficient model training. Additionally, the features ``ZipCodeOrigin'' and ``ZipMerchant'' are removed due to their lack of variability, and the remaining features are converted into numerical values to ensure data consistency.

The second dataset is a publicly available credit card fraud detection dataset \cite{ULB2013}, containing European cardholder transactions. Features are derived using Principal Component Analysis (PCA) on real user data. Each transaction is labeled as fraud (1) or non-fraud (0), with 28 PCA-transformed features, as well as ``Time'' and ``Amount''. 

\begin{itemize}
    \item \textbf{``Time''}: Indicates the elapsed time (in seconds) between transactions.
    \item \textbf{``Amount''}: Represents the monetary value of transactions.
    \item \textbf{``Class''}: The target variable, where 1 indicates fraud and 0 indicates non-fraud.
\end{itemize}

This dataset is unbalanced, containing 492 fraud cases out of 284,807 transactions, with the positive class representing only 0.172\% of all transactions. To address the class imbalance in both datasets, ``Random Under Sampling'' is implemented to create balanced datasets with a 50/50 ratio. Each dataset is reduced to 492 fraudulent and 492 non-fraudulent transactions. PCA is applied to the European dataset, reducing its dimensions to seven features named ``V1'' through ``V7''.

\begin{figure}[!h]
    \centering    
    \includegraphics[width=\linewidth]{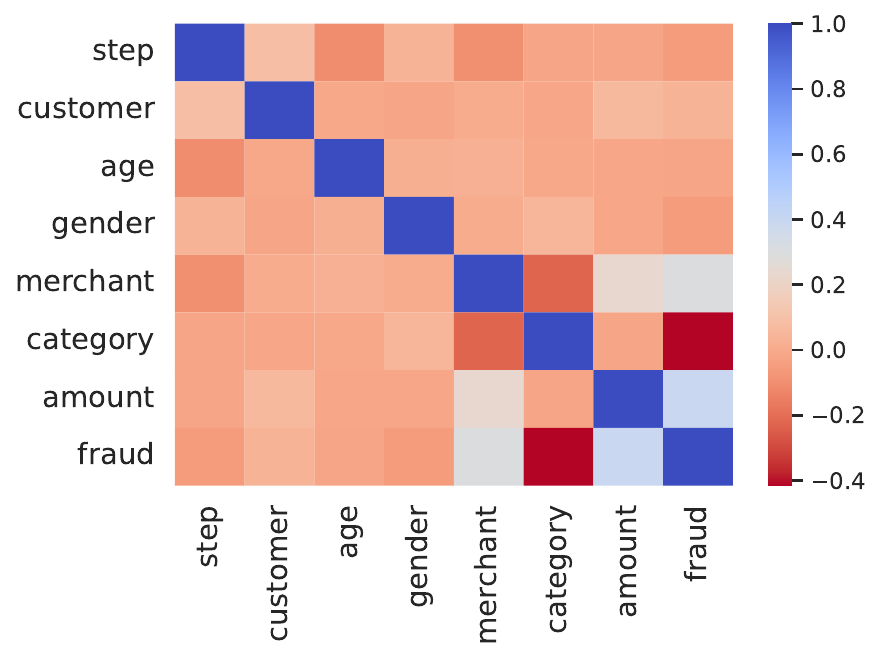}
    \vspace{-0.8cm}
    \caption{Correlation matrix of the BankSim dataset. Features such as ``Category'' show a strong negative correlation with the label ``Fraud'', while ``Merchant'' and ``Amount'' have positive correlations, indicating their importance in fraud detection.}
    \label{fig:Corr-Matrix-BankSim data}
\end{figure}

\begin{figure}[!h]
    \centering    
    \includegraphics[width=\linewidth]{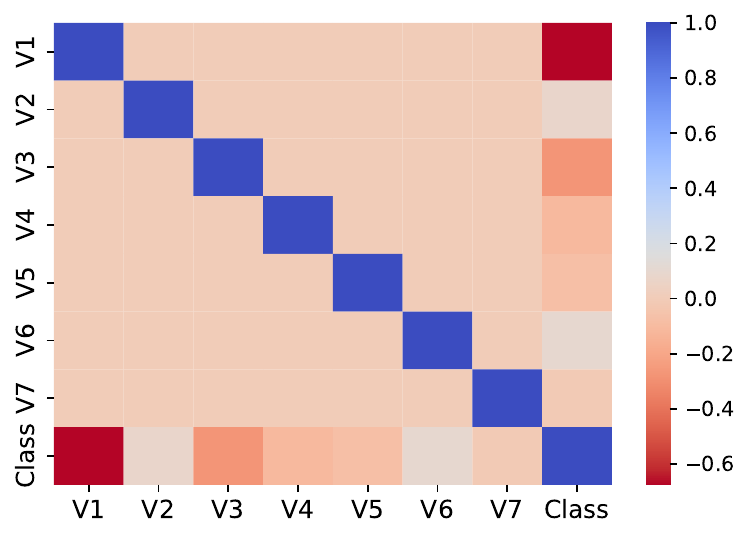}
    \vspace{-0.5cm}
    \caption{Correlation matrix of the European dataset. ``V1'' has a strong negative correlation with ``Class'', while ``V2'' and ``V6'' show moderate positive correlations, suggesting their relevance for fraud detection.}
    \label{fig:Corr-Matrix European Data}
\end{figure}

With the datasets balanced and preprocessed, visualizations are generated to understand feature relationships and their importance. Figs \ref{fig:Corr-Matrix-BankSim data} and \ref{fig:Corr-Matrix European Data} illustrate the correlation matrices, highlighting key interactions. For instance, in the BankSim dataset, ``Category'' has a strong negative correlation with ``Fraud'', while ``Merchant'' and ``Amount'' exhibit positive correlations. In the European dataset, features such as ``V1'' and ``V2'' show significant correlations with ``Class''.

\begin{figure}[!h]
    \centering
    \includegraphics[width=\linewidth]{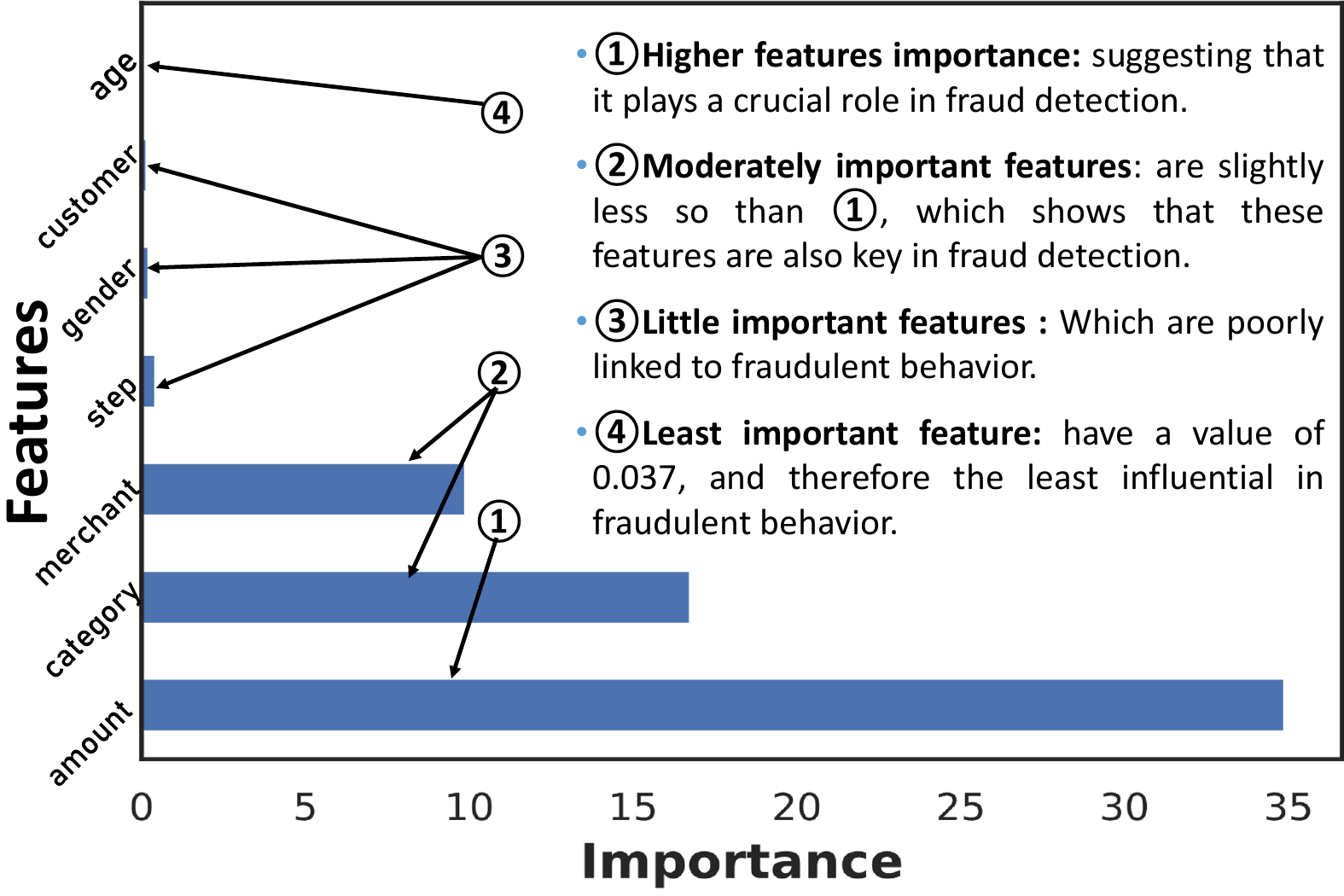}
    \vspace{-0.5cm}
    \caption{Feature importance for the BankSim dataset. ``Amount'', ``Category'', and ``Merchant'' are critical for fraud detection, while ``Step'', ``Gender'', and ``Age'' have minimal influence.}
    \label{fig:Feature-importance-Banksim data}
\end{figure}

\begin{figure}[!h]
    \centering
    \includegraphics[width=\linewidth]{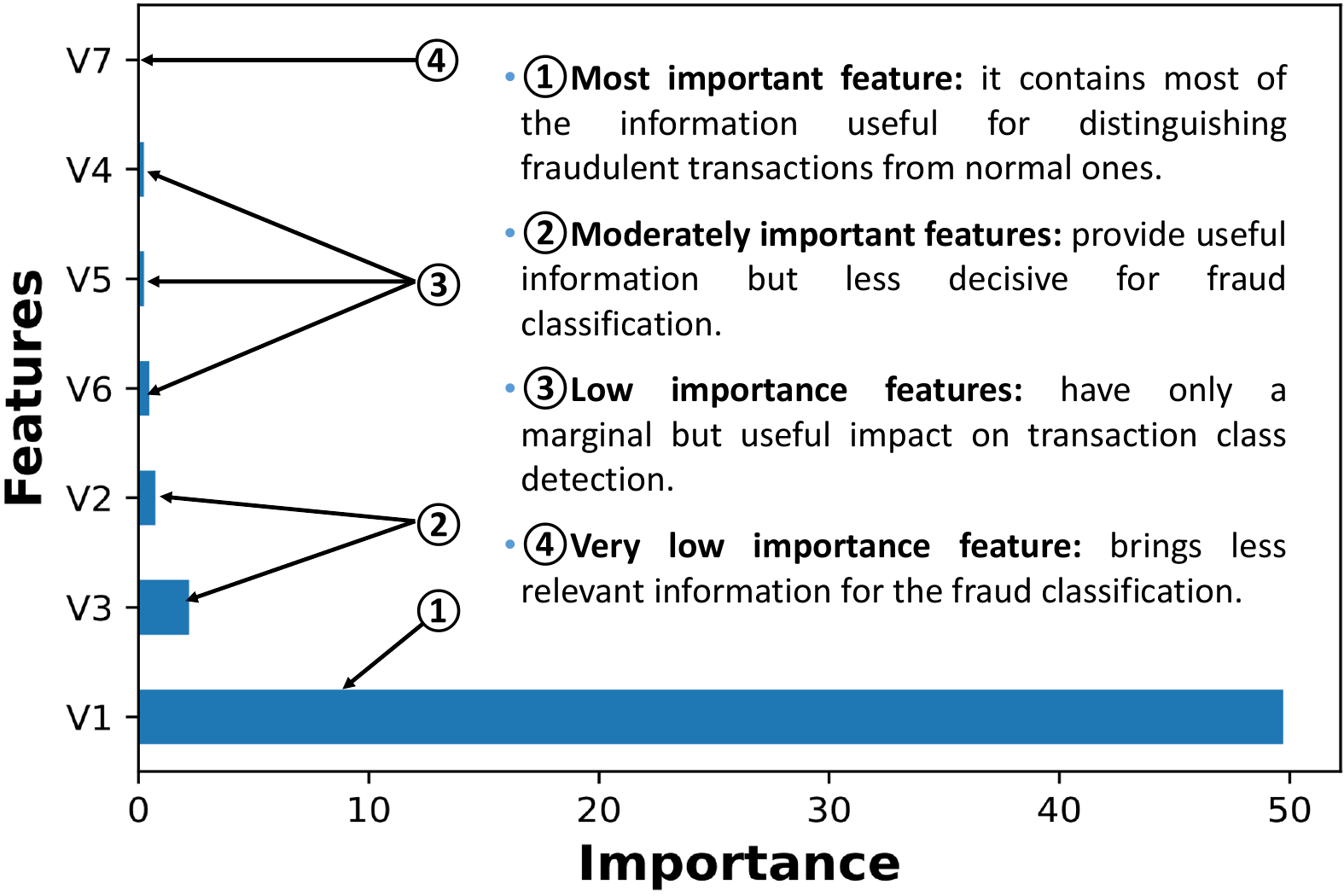}
    \vspace{-0.5cm}
    \caption{Feature importance for the European dataset. ``V1'' and ``V3'' are the most influential features for fraud prediction, with minimal contributions from other variables.}
    \label{fig:Feature-importance-European data}
\end{figure}

As shown in Figs. \ref{fig:Feature-importance-Banksim data} and \ref{fig:Feature-importance-European data}, the analysis identifies specific features as particularly influential for predicting fraud. This insight allows us to focus on these key features to enhance the model's performance and accuracy.

\subsection{Experimental Setup}

To ensure the reliability of the fraud detection analysis, a comprehensive data preparation process is conducted to confirm the datasets' quality and suitability for training and evaluating robust models. The datasets are split into 80\% training and 20\% testing subsets. The training set denoted as $X_{\text{train}}$ and $y_{\text{train}}$, is used to train the models, while the testing set, represented by $X_{\text{test}}$ and $y_{\text{test}}$, remains unseen during training and is used to evaluate model performance.

For each dataset, the feature matrix $X$ includes all relevant features except the label column, which serves as the target variable $y$. The target variable distinguishes between fraudulent transactions (coded as ``1'') and non-fraudulent transactions (coded as ``0''), enabling the models to learn patterns and accurately classify new data. To normalize the feature values and enhance model stability, scikit-learn's \texttt{MinMaxScaler} is employed, scaling all feature values to the range [0, 1] without altering their distribution. Each QML model is tested under identical configurations, with three feature map architectures and four ansatz architectures, resulting in twelve unique configurations per model. The analysis is carried out on two datasets.

Model optimization is handled using Qiskit's COBYLA optimizer, chosen for its efficiency and ability to converge to optimal solutions effectively. The Aer backend with the \texttt{QasmSimulator} is utilized to simulate quantum circuits, ensuring a seamless and transparent training process. Following training, the models' performance is meticulously evaluated using key metrics, including accuracy, precision, recall, F1-score, and loss.

The hyperparameters and configurations used in this study are summarized in Table \ref{tab:parameter_values}. These configurations are carefully selected to ensure an optimal balance between computational efficiency and model performance.

\begin{table}[h]
\caption{Configuration and hyperparameters used.}
    \centering
    \begin{tabular}{cc}
        \toprule
        \textbf{Parameter} & \textbf{Value}\\
        \midrule
        Number of Qubits & 4, 6, \textbf{7}\\
        Optimizer & \textbf{COBYLA}, ADAM, Gradient Descent\\
        Loss function & Cross-Entropy\\
        Maximum Iterations & 350\\
        Test Set Size & 20\%\\
        Performance Metrics & Accuracy, Precision, Recall, F1-score, Loss\\
        Repetition of Feature Map & \textbf{1}, 2\\
        Repetition of Ansatz & \textbf{1}, 2\\
        Libraries and Backend & Qiskit, QasmSimulator\\
        \bottomrule
    \end{tabular}
    \label{tab:parameter_values}
\end{table}

\subsection{BankSim Dataset}
\begin{figure}[htpb]
    \centering
    \includegraphics[width=\linewidth]{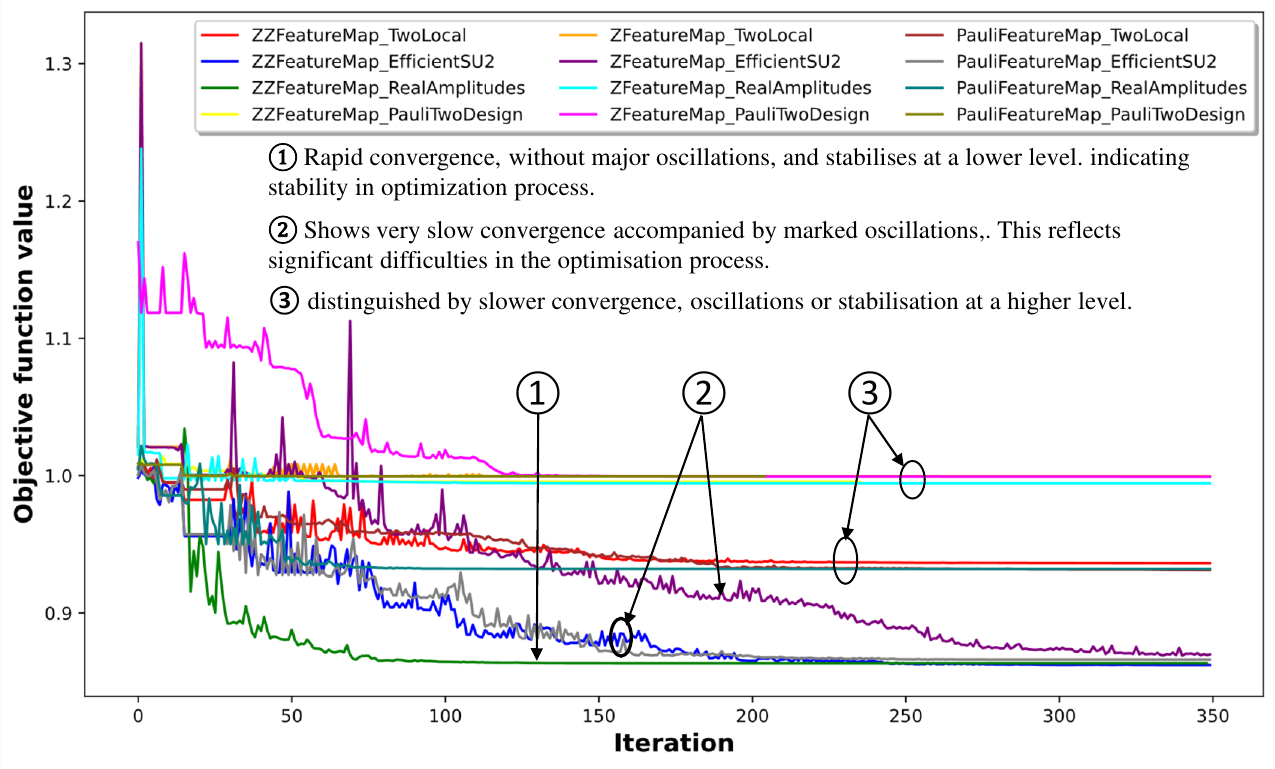}
    \caption{VQC loss function on the BankSim dataset. Most configurations demonstrate stabilization over time, indicating consistent optimization behavior. However, some configurations exhibit a zigzag pattern during convergence, reflecting oscillatory dynamics before eventual stabilization. 
    }
    \label{fig:Loss QVC1}
\end{figure}

\begin{figure}[htpb]
    \centering
    \includegraphics[width=\linewidth]{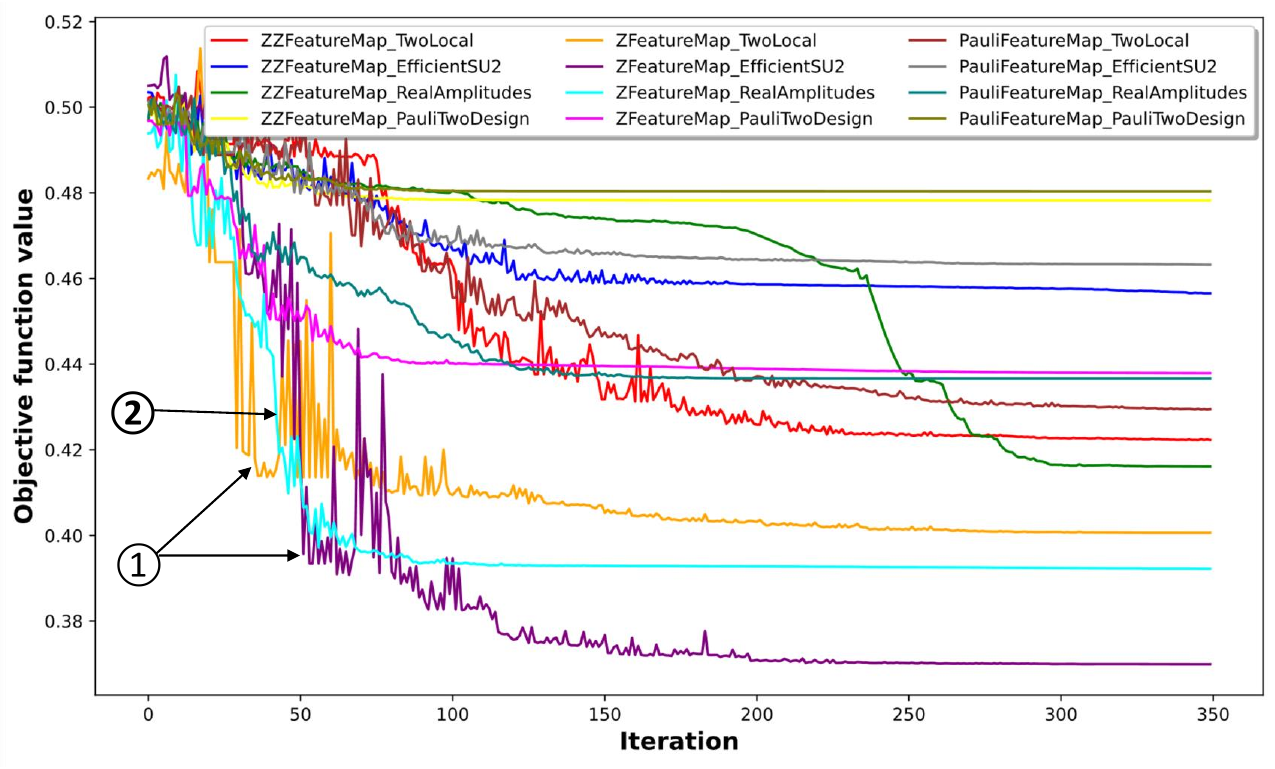}
    \caption{SQNN loss function on the BankSim dataset. The configuration in \textcircled{\raisebox{-0.9pt}{1}} achieves convergence with minimal loss but does not correspond to the absolute minimum loss. Configurations labeled as \textcircled{\raisebox{-0.9pt}{2}} exhibit similar behaviors across different levels of loss, characterized by a zigzag pattern during the initial optimization phase, eventually stabilizing after requiring more iterations to converge.}
    \label{fig:Loss SQNN1}
\end{figure}

\begin{figure}[htpb]
    \centering
    \includegraphics[width=\linewidth]{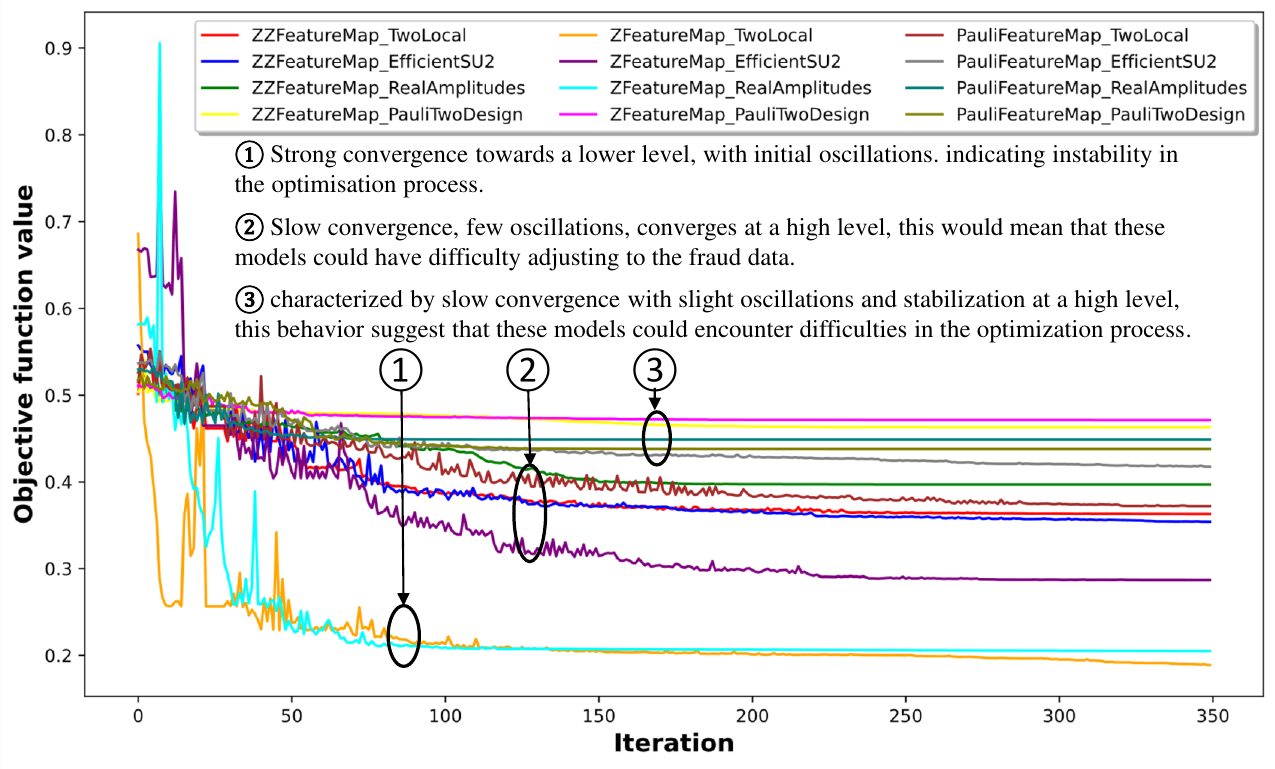}
    \caption{EQNN loss function on the BankSim dataset. Most configurations demonstrate steady convergence, stabilizing after a period of oscillations and reaching varying levels of loss. While some achieve minimal loss, others stabilize at higher values, reflecting differences in optimization effectiveness.}
    \label{fig:Loss EQNN2.pdf}
\end{figure}

Table \ref{tab:table VQC1} highlights the performance of the VQC model. Notably, the ZZ and Pauli feature maps paired with the Efficient SU2 ansatz yield the highest F1-scores of 0.68 and 0.71, respectively. These results suggest that introducing entanglement through the feature map improves the VQC's ability to learn complex interactions. Fig. \ref{fig:Loss QVC1} supports this, as most configurations stabilize after oscillatory dynamics, indicating successful optimization.
Similarly, Table \ref{tab:table SQNN1} reveals the robust performance of the SQNN model. The Pauli feature map combined with Two Local and Real Amplitudes achieves F1-scores of 0.84 and 0.85, respectively, showcasing the effectiveness of these configurations. While Fig. \ref{fig:Loss SQNN1} demonstrates consistent convergence behavior across these configurations, with some achieving minimal loss values after initial oscillations.
In contrast, Table \ref{tab:table EQNN1} shows that the EQNN model generally underperforms compared to VQC and SQNN. While the Pauli feature map with Pauli Two Design achieves a peak F1-score of 0.59, configurations involving the Z feature map yield significantly lower scores, underscoring the importance of entanglement in feature maps. Fig. \ref{fig:Loss EQNN2.pdf} corroborates these findings, representing varying levels of convergence stability across configurations.

\begin{table}[htpb]
\caption{Performance metrics of the VQC model on the BankSim dataset.}
\centering
\setlength{\tabcolsep}{1pt} 
\begin{tabular}{ccccccc}
\toprule
\textbf{Model} & \textbf{Feature Map} & \textbf{Ansatz} & \textbf{Accuracy} & \textbf{Precision} & \textbf{Recall} & \textbf{F1\_score} \\ \bottomrule
     & Z & Real Amplitudes & 0.52 & 0.48 & 0.8 & 0.6 \\ 
     &             &\textbf{Two Local}       & \textbf{0.54} &\textbf{0.5} &\textbf{0.84} &\textbf{0.62} \\ 
     &             & Efficient SU2   & 0.65 & 0.68 & 0.44 & 0.54 \\ 
     &             & Pauli Two Design & 0.55 & 0.51 & 0.51 & 0.51 \\ 
VQC  & ZZ & Real Amplitudes & 0.74 & 0.75 & 0.62 & 0.68 \\ 
     &             &\textbf{Two Local}       &\textbf{0.65} &\textbf{0.62} &\textbf{0.6} &\textbf{0.61} \\ 
     &             & Efficient SU2   & 0.76 & 0.85 & 0.56 & 0.68 \\ 
     &             & Pauli Two Design & 0.51 & 0.46 & 0.42 & 0.44 \\ 
     & Pauli& Real Amplitudes & 0.73 & 0.69 & 0.72 & 0.7 \\ 
     &             &\textbf{Two Local}       &\textbf{0.73} &\textbf{0.69} &\textbf{0.72} &\textbf{0.71}\\ 
     &             & Efficient SU2   & 0.76 & 0.85 & 0.56 & 0.68 \\ 
     &             & Pauli Two Design & 0.5 & 0.45 & 0.46 & 0.46 \\
     \bottomrule
\end{tabular}
\label{tab:table VQC1}
\end{table}

\begin{table}[htpb]
\caption{Performance metrics of the SQNN model on the BankSim dataset.}
\centering
\setlength{\tabcolsep}{1pt} 
\begin{tabular}{ccccccc}
\toprule
\textbf{Model} & \textbf{Feature Map} & \textbf{Ansatz} & \textbf{Accuracy} & \textbf{Precision} & \textbf{Recall} & \textbf{F1\_score} \\ \bottomrule
     & Z &\textbf{Real Amplitudes} &\textbf{0.86} &\textbf{0.82} &\textbf{0.88} &\textbf{0.85} \\ 
     &             & Two Local       & 0.73 & 0.75 & 0.58 & 0.66 \\ 
     &             & Efficient SU2   & 0.83 & 0.81 & 0.82 & 0.81 \\ 
     &             & Pauli Two Design & 0.85 & 0.78 & 0.91 & 0.84 \\ 
SQNN & ZZ & Real Amplitudes & 0.74 & 0.77 & 0.61 & 0.68 \\ 
     &             &\textbf{Two Local}       &\textbf{0.83} &\textbf{0.85} &\textbf{0.76} &\textbf{0.8} \\ 
     &             & Efficient SU2   & 0.77 & 0.73 & 0.76 & 0.75 \\ 
     &             & Pauli Two Design & 0.64 & 0.63 & 0.51 & 0.56 \\ 
     & Pauli & Real Amplitudes & 0.82 & 0.82 & 0.76 & 0.79 \\ 
     &             &\textbf{Two Local}       & \textbf{0.86} &\textbf{0.84} &\textbf{0.84} &\textbf{0.84} \\ 
     &             & Efficient SU2   & 0.77 & 0.81 & 0.64 & 0.72 \\ 
     &             & Pauli Two Design & 0.64 & 0.59 & 0.7 & 0.64 \\
     \bottomrule
\end{tabular}
\label{tab:table SQNN1}
\end{table}

\begin{table}[h]
\caption{Performance metrics of the EQNN model on the BankSim dataset.}
\centering
\setlength{\tabcolsep}{1pt} 
\begin{tabular}{ccccccc}
\toprule
\textbf{Model} & \textbf{Feature Map} & \textbf{Ansatz} & \textbf{Accuracy} & \textbf{Precision} & \textbf{Recall} & \textbf{F1\_score} \\ \bottomrule
     & Z& Real Amplitudes & 0.45 & 0.22 & 0.49 & 0.31 \\ 
     &             & Two Local       & 0.46 & 0.22 & 0.50 & 0.31 \\ 
     &             & Efficient SU2   & 0.46 & 0.22 & 0.50 & 0.31 \\ 
     &             & Pauli Two Design & 0.32 & 0.28 & 0.34 & 0.28 \\ 
EQNN & ZZ & Real Amplitudes & 0.45 & 0.46 & 0.48 & 0.37 \\ 
     &             & Two Local       & 0.50 & 0.57 & 0.53 & 0.42 \\
     &             & Efficient SU2   & 0.53 & 0.60 & 0.56 & 0.49 \\ 
     &             &\textbf{Pauli Two Design} &\textbf{0.51} &\textbf{0.52} &\textbf{0.52} &\textbf{0.50} \\ 
     & Pauli & Real Amplitudes & 0.47 & 0.49 & 0.49 & 0.41 \\ 
     &             & Two Local       & 0.55 & 0.63 & 0.58 & 0.51 \\ 
     &             & Efficient SU2   & 0.52 & 0.53 & 0.52 & 0.51 \\ 
     &             &\textbf{Pauli Two Design} &\textbf{0.60} &\textbf{0.61} &\textbf{0.61} &\textbf{0.59} \\ 
     \bottomrule
\end{tabular}
\label{tab:table EQNN1}
\end{table}

\subsubsection{Analysis of Feature Maps}
Feature maps play a crucial role in capturing complex data patterns. The Z feature map, while computationally efficient, lacks expressivity due to the absence of entanglement. This limitation is evident in the EQNN model's F1-scores, which stagnate at 0.31 across most ansatz configurations (Table \ref{tab:table EQNN1}). Conversely, the ZZ feature map introduces pairwise entanglement, significantly enhancing performance for VQC and SQNN. For instance, the ZZ feature map paired with Efficient SU2 achieves F1-scores of 0.68 and 0.75 for VQC and SQNN, respectively.
The Pauli feature map emerges as the most robust, consistently delivering superior results across all models. In SQNN, its configuration with Two Local achieves the highest F1-score of 0.84, as shown in Table \ref{tab:table SQNN1}. For EQNN, the Pauli feature map compensates for architectural limitations, achieving a peak F1-score of 0.59 with Pauli Two Design. These findings emphasize that feature maps incorporating entanglement, such as ZZ and Pauli, enhance model performance, especially when paired with expressive ansatz configurations.
\subsubsection{Analysis of Ansatz}
The choice of ansatz significantly impacts the quantum circuit's ability to model complex functions. Real Amplitudes, with its simple configuration, performs competitively in SQNN, achieving an F1-score of 0.85 with the Z feature map (Table \ref{tab:table SQNN1}). However, it struggles to capture intricate interactions in EQNN, resulting in suboptimal performance.

Two Local, characterized by alternating single-qubit rotations and entangling gates, exhibits strong performance in SQNN (F1-score of 0.84 with the Pauli feature map). However, its performance in EQNN remains moderate, likely due to limited parametrization. Efficient SU2 emerges as the most versatile ansatz, achieving consistently strong metrics across models. Notably, it delivers F1-scores of 0.68 and 0.81 for VQC and SQNN, respectively, when paired with ZZ and Z feature maps.

In contrast, Pauli Two Design shows inconsistent results. While it achieves its best performance in EQNN with the Pauli feature map (F1-score of 0.59), its performance across other configurations varies significantly. This underscores the need to carefully align ansatz configurations with the chosen feature map and model architecture to maximize performance.
\subsection{European Dataset}
\begin{figure}[htpb]
    \centering    \includegraphics[width=\linewidth]{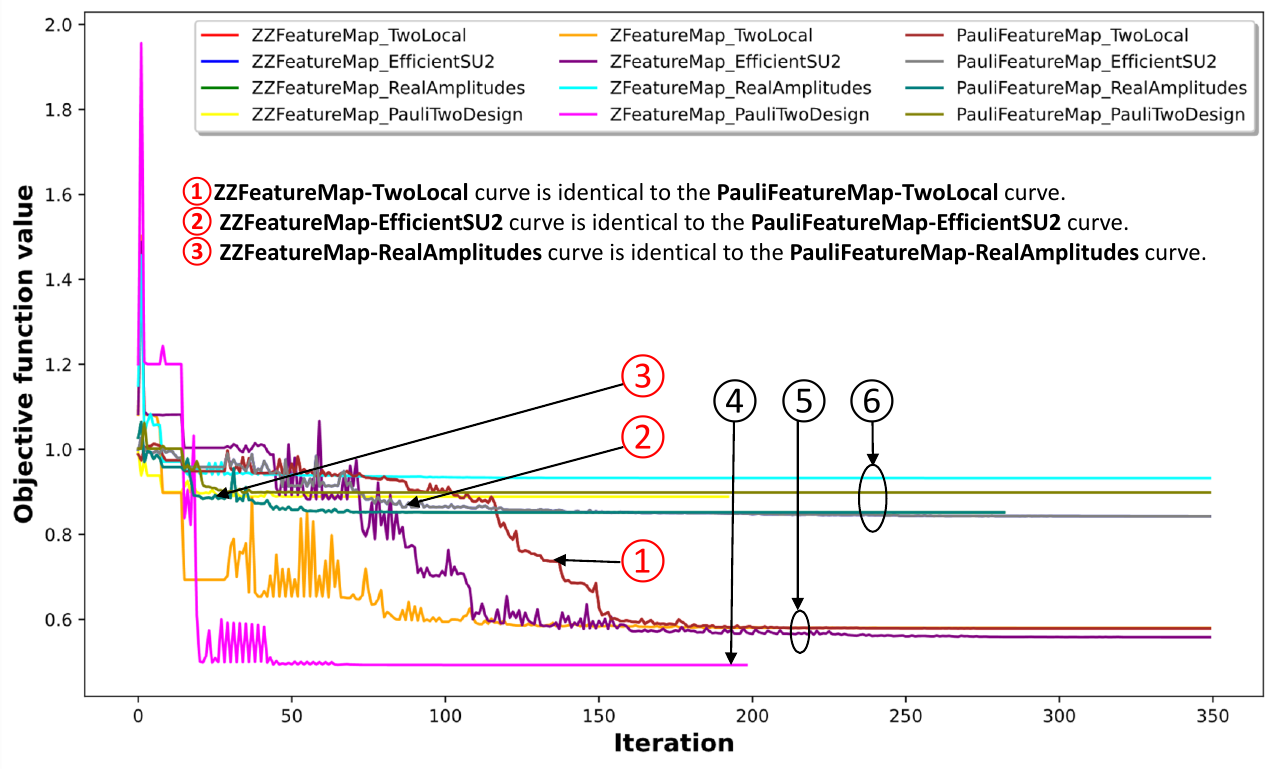}
    \caption{VQC loss function on the European dataset. The configuration in \textcircled{\raisebox{-0.9pt}{4}} demonstrates fast convergence with minimal loss, showing some oscillations during the first 50 iterations before stabilizing. The configurations in \textcircled{\raisebox{-0.9pt}{5}} exhibit similar behavior, converging steadily over 200 iterations to achieve relatively low loss values. In contrast, the configurations in \textcircled{\raisebox{-0.9pt}{6}} converge more rapidly but stabilize at a higher loss value, indicating suboptimal performance.}
    \label{fig:Loss QVC2}
\end{figure}

\begin{figure}[htpb]
    \centering
    \includegraphics[width=\linewidth]{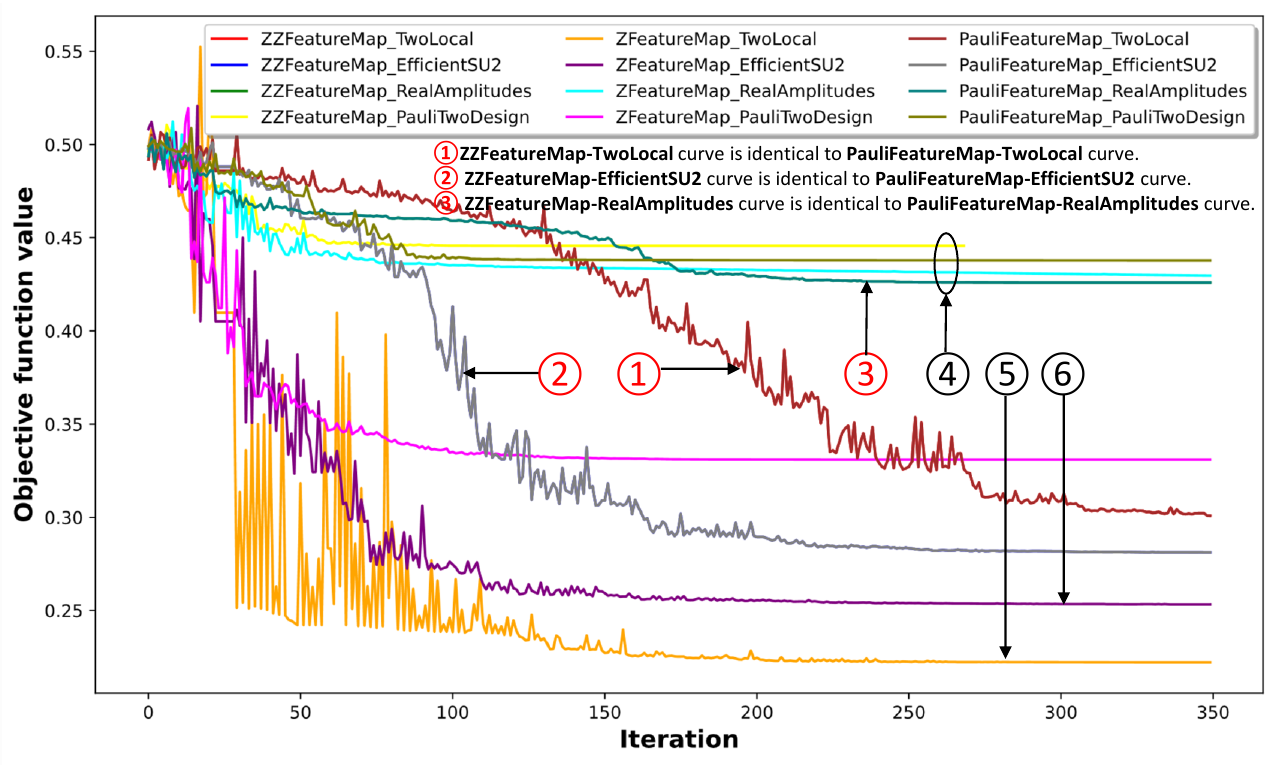}
    \caption{SQNN loss function on the European dataset. The configurations highlighted in \textcircled{\raisebox{-0.9pt}{4}} demonstrate faster convergence but do not reach the lowest loss value. The configuration in \textcircled{\raisebox{-0.9pt}{5}} converges quickly and achieves the minimum loss among all configurations, making it the most favorable. In contrast, the configuration marked as \textcircled{\raisebox{-0.9pt}{6}} stabilizes first but achieves a medium loss value compared to the others, indicating limited performance.}
    \label{fig:Loss SQNN2}
\end{figure}

\begin{figure}[htpb]
    \centering
    \includegraphics[width=\linewidth]{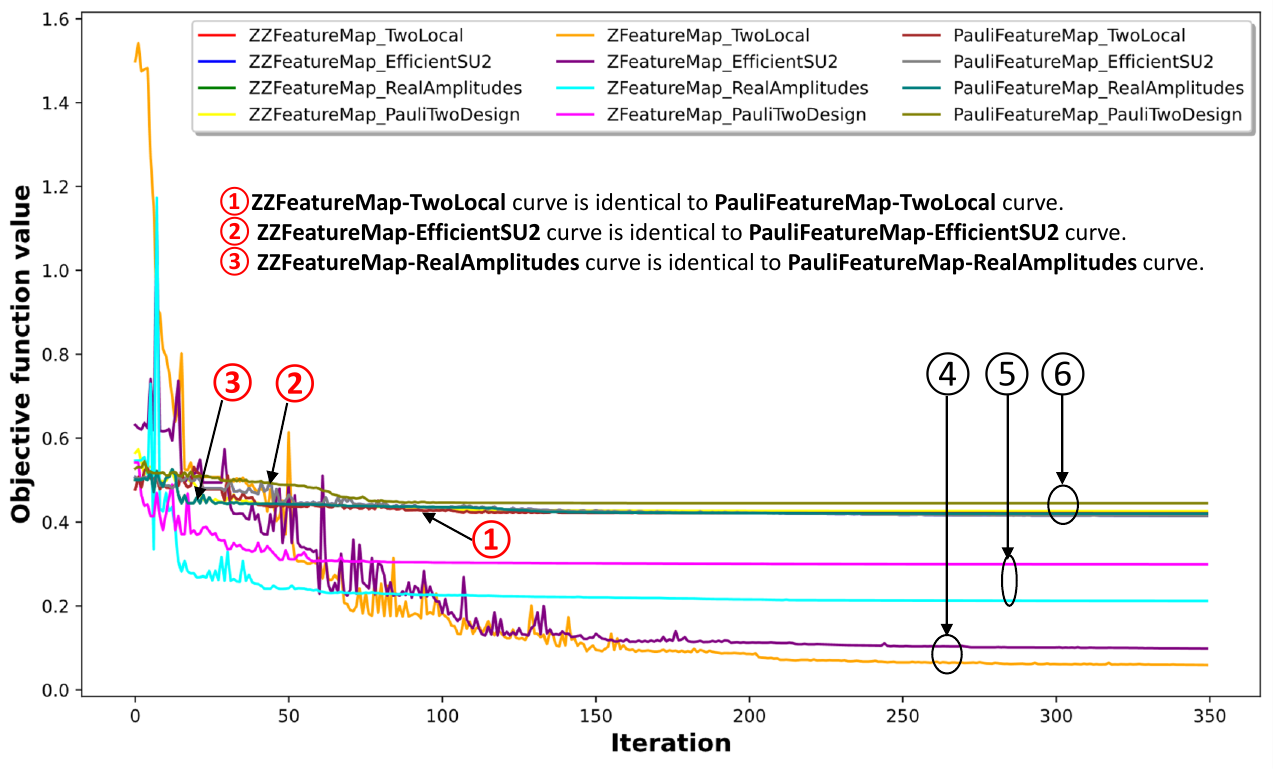}
    \caption{EQNN loss function on the European dataset. The configurations highlighted in \textcircled{\raisebox{-0.9pt}{4}} exhibit slow convergence to the lowest loss value, accompanied by oscillatory behavior. In contrast, the configurations marked as \textcircled{\raisebox{-0.9pt}{5}} demonstrate slower convergence and stabilize at a higher loss value, suggesting reduced effectiveness. Finally, the configurations labeled \textcircled{\raisebox{-0.9pt}{6}} achieve faster convergence, indicating a more favorable performance among the analyzed configurations.}
    \label{fig:Loss_EQNN2.pdf}
\end{figure}

The VQC model's results, as presented in Table \ref{tab:table VQC2}, show that the configuration of the Z feature map with the Pauli Two Design ansatz achieves the highest F1-score of 0.88. Similarly, the ZZ and Pauli feature maps paired with Two Local yield strong F1-scores of 0.83 each, indicating the robustness of Two Local across feature maps. Fig. \ref{fig:Loss QVC2} further supports these findings, highlighting configurations with steady convergence and minimal loss.
Table \ref{tab:table SQNN2} demonstrates that SQNN achieves its best performance when the Z feature map is paired with the Two Local ansatz (F1-score of 0.80). The configuration of the Pauli feature map with the Two Local ansatz also performs well, achieving an F1-score of 0.81. These configurations exhibit consistent convergence behavior, as shown in Fig. \ref{fig:Loss SQNN2}. In contrast, other configurations either stabilize at higher loss values or demonstrate slower convergence.
The EQNN model's performance, as outlined in Table \ref{tab:table EQNN2}, generally lags behind VQC and SQNN. The best performance is achieved with the ZZ feature map and the Two Local ansatz, yielding an F1-score of 0.52. Fig. \ref{fig:Loss_EQNN2.pdf} illustrates the optimization dynamics, where most of these configurations exhibit oscillatory behavior before stabilizing at suboptimal loss values.

\begin{table}[htpb]
\caption{Performance metrics of the VQC model on the European dataset.}
\centering
\setlength{\tabcolsep}{1pt} 
\begin{tabular}{ccccccc}
\toprule
\textbf{Model} & \textbf{Feature Map} & \textbf{Ansatz} & \textbf{Accuracy} & \textbf{Precision} & \textbf{Recall} & \textbf{F1\_score} \\ \bottomrule
     & Z & Real Amplitudes & 0.7 & 1 & 0.34 & 0.51 \\ 
     &             & Two Local       & 0.88 & 1 & 0.73 & 0.84 \\ 
     &             & Efficient SU2   & 0.86 & 1 & 0.68 & 0.81 \\ 
     &             &\textbf{Pauli Two Design} &\textbf{0.9} &\textbf{1} &\textbf{0.78} &\textbf{0.88} \\
VQC & ZZ & Real Amplitudes & 0.74 & 1 & 0.43 & 0.6 \\ 
     &             &\textbf{Two Local}       &\textbf{0.87} &\textbf{1} &\textbf{0.71} &\textbf{0.83} \\ 
     &             & Efficient SU2   & 0.78 & 0.94 & 0.55 & 0.69 \\ 
     &             & Pauli Two Design & 0.73 & 0.77 & 0.56 & 0.65 \\ 
     & Pauli & Real Amplitudes & 0.74 & 1 & 0.43 & 0.6 \\ 
     &             &\textbf{Two Local}       &\textbf{0.87} &\textbf{1} &\textbf{0.71} &\textbf{0.83} \\ 
     &             & Efficient SU2   & 0.78 & 0.94 & 0.55 & 0.69 \\ 
     &             & Pauli Two Design & 0.81 & 0.96 & 0.6 & 0.73 \\
     \bottomrule
\end{tabular}
\label{tab:table VQC2}
\end{table}
\begin{table}[h]
\caption{Performance metrics of the SQNN model on the European dataset.}
\centering
\setlength{\tabcolsep}{1pt} 
\begin{tabular}{ccccccc}
\toprule
\textbf{Model} & \textbf{Feature Map} & \textbf{Ansatz} & \textbf{Accuracy} & \textbf{Precision} & \textbf{Recall} & \textbf{F1\_score} \\ \bottomrule
     & Z & Real Amplitudes & 0.64 & 1 & 0.21 & 0.34 \\ 
     &             &\textbf{Two Local}       &\textbf{0.85} &\textbf{1} &\textbf{0.67} &\textbf{0.8} \\ 
     &             & Efficient SU2   & 0.8 & 1 & 0.56 & 0.72 \\ 
     &             & Pauli Two Design & 0.75 & 1 & 0.45 & 0.62 \\ 
SQNN & ZZ & Real Amplitudes & 0.65 & 1 & 0.23 & 0.37 \\ 
     &             &\textbf{Two Local}       &\textbf{0.86}  &\textbf{1} &\textbf{0.68} &\textbf{0.68} \\ 
     &             & Efficient SU2   & 0.8 & 1 & 0.56 & 0.72 \\ 
     &             & Pauli Two Design & 0.69 & 0.89 & 0.36 & 0.51 \\ 
     & Pauli & Real Amplitudes & 0.65 & 1 & 0.23 & 0.37 \\ 
     &             &\textbf{Two Local}       &\textbf{0.86} &\textbf{1} &\textbf{0.68} &\textbf{0.81} \\ 
     &             & Efficient SU2   & 0.8 & 1 & 0.56 & 0.72 \\ 
     &             & Pauli Two Design & 0.8 & 0.94 & 0.6 & 0.73 \\
     \bottomrule
\end{tabular}
\label{tab:table SQNN2}
\end{table}
\begin{table}[h]
\centering
\caption{EQNN model performance metrics of the SQNN model on the European dataset.}
\setlength{\tabcolsep}{1pt} 
\begin{tabular}{ccccccc}
\toprule
\textbf{Model} & \textbf{Feature Map} & \textbf{Ansatz} & \textbf{Accuracy} & \textbf{Precision} & \textbf{Recall} & \textbf{F1\_score} \\ \bottomrule
     & Z & Real Amplitudes & 0.45 & 0.22 & 0.50 & 0.31 \\ 
     &             &\textbf{Two Local}       &\textbf{0.49} &\textbf{0.73} &\textbf{0.52} &\textbf{0.37} \\ 
     &             & Efficient SU2   & 0.46 & 0.22 & 0.5 & 0.31 \\ 
     &             & Pauli Two Design & 0.41 & 0.21 & 0.45 & 0.29 \\
EQNN & ZZ & Real Amplitudes & 0.45 & 0.45 & 0.47 & 0.40 \\ 
     &             &\textbf{Two Local}       &\textbf{0.55} &\textbf{0.61} &\textbf{0.57} &\textbf{0.52} \\ 
     &             & Efficient SU2   & 0.45 & 0.45 & 0.47 & 0.41 \\ 
     &             & Pauli Two Design & 0.44 & 0.43 & 0.46 & 0.39 \\ 
     & Pauli & Real Amplitudes & 0.45 & 0.45 & 0.47 & 0.40 \\ 
     &             &\textbf{Two Local}       &\textbf{0.55} &\textbf{0.61} &\textbf{0.57} &\textbf{0.52} \\ 
     &             & Efficient SU2   & 0.45 & 0.45 & 0.47 & 0.41 \\ 
     &             & Pauli Two Design & 0.47 & 0.48 & 0.48 & 0.44 \\ 
     \bottomrule
\end{tabular}
\label{tab:table EQNN2}
\end{table}

\subsubsection{Analysis of Feature Maps}
The Z feature map consistently delivers strong results across models. For instance, in VQC, its configuration with the Pauli Two Design ansatz achieves the highest F1-score of 0.88 (Table \ref{tab:table VQC2}). Similarly, in SQNN, the Z feature map paired with the Two Local ansatz yields an F1-score of 0.80 (Table \ref{tab:table SQNN2}). However, the Z feature map's performance in EQNN is limited, with the highest F1-score being 0.52.
The ZZ feature map introduces pairwise entanglement, leading to competitive results. In VQC, its configuration with the Two Local ansatz achieves an F1-score of 0.83. Similarly, in SQNN, the ZZ feature map paired with the Efficient SU2 ansatz results in an F1-score of 0.72. However, its performance in EQNN remains limited, with an F1-score of 0.52 when paired with the Two Local ansatz.
The Pauli feature map demonstrates consistent performance across VQC and SQNN. In VQC, its configuration with the Two Local ansatz achieves an F1-score of 0.83, while in SQNN, the same configuration results in an F1-score of 0.81. However, in EQNN, its best performance remains low, with an F1-score of 0.52.

\subsubsection{Analysis of Ansatz}
 Real Amplitudes, while computationally efficient, struggle with complex interactions. For example, in VQC, it achieves an F1-score of 0.6 with the Pauli feature map, but its performance in SQNN and EQNN remains limited.
Two Local emerges as the most robust ansatz, consistently delivering strong results across models. In VQC, it achieves F1-scores of 0.84 and 0.83 with the ZZ and Pauli feature maps, respectively. In SQNN, it attains an F1-score of 0.81 with the Pauli feature map. This versatility underscores its ability to effectively capture intricate data patterns when paired with expressive feature maps.
Efficient SU2 shows strong performance in VQC, achieving an F1-score of 0.81 with the Z feature map. However, its performance in EQNN is moderate, likely due to insufficient parametrization.
Pauli Two Design exhibits erratic outcomes. In VQC, it achieves an F1-score of 0.88 with the Z feature map, but its performance in SQNN and EQNN is inconsistent, with F1-scores of 0.73 and 0.44, respectively. This variability suggests that Pauli Two Design's effectiveness is highly dependent on the model architecture and feature map used.
\subsection{Statistical Analysis}
To confirm the validity of the differences observed between the QML model architectures, we perform a one-way ANOVA test on the F1-score.  ANOVA evaluates whether the means of multiple groups differ significantly by comparing the variance between groups to the variance within groups, expressed as an F-value. A larger F-value and a smaller p-value ($p < 0.05$) indicate that the differences are unlikely due to random chance. This analysis considers four main experimental factors: the model, the feature map, the ansatz, and the dataset.
\begin{table}[htpb]
\centering
\caption{Results of the one-way ANOVA test for the F1-score.}
\begin{tabular}{cccc}
\toprule
\textbf{Factor}    & \textbf{F}     & \textbf{p-value}      & \textbf{Interpretation}         \\
\midrule
Dataset             & 1.71           & 0.1917                &  Not significant                \\
Model               & 335.08         & $1.18 \times 10^{-102}$ & Highly significant           \\
Feature Map          & 8.96           & $1.44 \times 10^{-4}$  & Significant                    \\
Ansatz              & 10.96          & $7.51 \times 10^{-23}$ & highly significant          \\
\bottomrule
\end{tabular}
\label{tab-anova}
\end{table}

As shown in Table~\ref{tab-anova}, the results reveal that the choice of QML model architecture has a statistically significant effect on F1-score performance. Specifically, the ``Model'' factor shows a very high F-value (335.08) and an extremely low p-value ($1.18 \times 10^{-102}$), indicating a highly significant influence. The feature map and ansatz factors also yield significant F-values (8.96 and 10.96, respectively), with corresponding p-values well below 0.05, confirming their impact. In contrast, the Dataset factor is not significant (F = 1.71, $p = 0.1917$), suggesting consistent performance across datasets.

\begin{figure}[htpb]
    \centering
    \includegraphics[width=\linewidth]{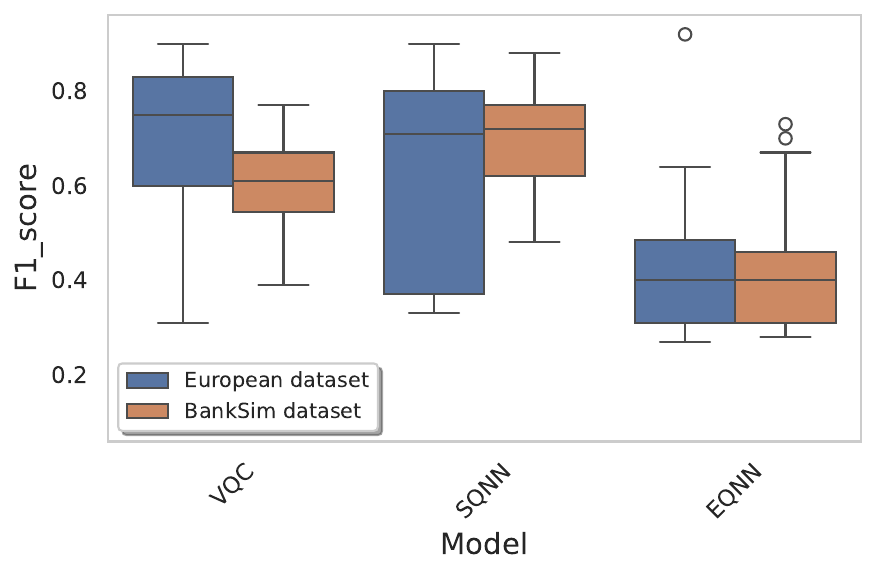}
    \caption{Boxplot shows the distribution of the F1-scores obtained by the three QML models evaluated on the BankSim and European datasets. Each box shows the distribution of F1-scores across ten repeated runs for each model.}
    \label{boxplot}
\end{figure}

To support these results, Fig.~\ref{boxplot} presents the distribution of F1-scores for each model, distinguishing between the datasets used.
These graphical results show that the VQC model performs best on the European dataset, with a median F1-score of around 0.75 and moderate variance, which aligns with our results. This suggests that the VQC model can effectively capture the characteristic patterns of real fraudulent transactions. On the BankSim dataset, its performance is slightly lower, with a median of around 0.65 and greater variance, possibly reflecting sensitivity to the synthetic nature of the data.

The SQNN model performs remarkably well and consistently across both datasets, with a median close to 0.8 on BankSim. However, the results on the European dataset show greater variability, which could indicate increased sensitivity to specific architectural combinations. The stability observed on BankSim, in contrast, suggests better generalization.
The EQNN model performs significantly worse, with a median F1-score of around 0.4 on both datasets. The low dispersion observed, combined with a few outliers, suggests limited adaptability, which reduces its effectiveness in complex financial fraud detection scenarios.

This statistical analysis reinforces the validity of our comparative study and highlights the critical importance of QML model selection in financial fraud detection. It also underscores the need to systematically explore different combinations of feature maps and ansatz within quantum model architectures. Our conclusions are thus based not only on empirical performance averages but also on statistically significant differences confirmed through rigorous analysis of variance.

\subsection{Noise Analysis}
\begin{figure}[htpb]
 \centering
 \includegraphics[width=\linewidth]{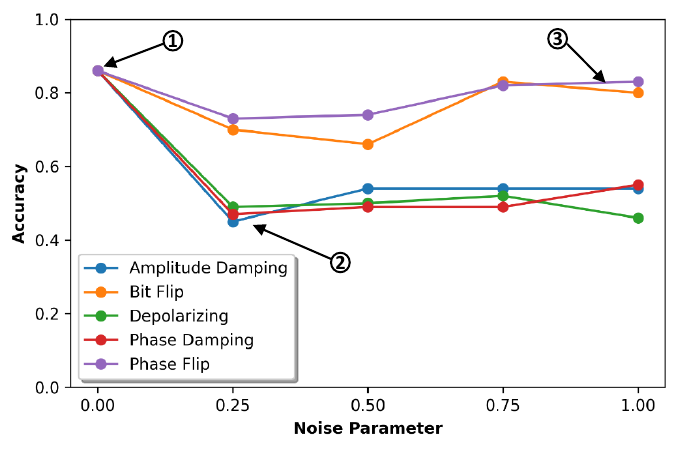}
 \caption{Impact of quantum noise on the SQNN model on the BankSim dataset.}
 \label{SQNN_0}
 \end{figure}

\begin{figure}[htpb]
 \centering
 \includegraphics[width=\linewidth]{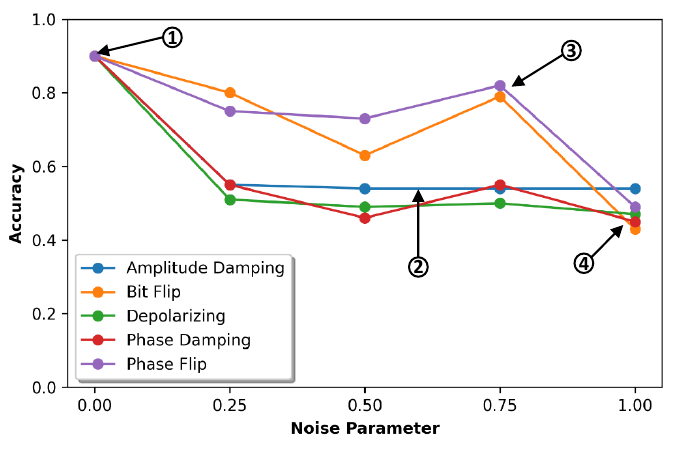}
 \caption{Impact of quantum noise on the VQC model on the European dataset.}
 \label{VQC Noise}
 \end{figure}

In this study, we evaluate the robustness of only the best-performing model selected for each dataset under realistic quantum noise conditions, simulating the behavior of near-term quantum devices. We consider five representative noise channels, amplitude damping, bit flip, depolarizing, phase damping, and phase flip, each implemented using Kraus operators \cite{innan2025optimizing}. The SQNN model is selected for the BankSim dataset, and the VQC model is chosen for the European dataset, based on their superior performance in noise-free settings. We assess how each model responds to increasing levels of quantum noise, providing insights into their stability and resilience across different interference scenarios.
Starting with the SQNN model on the BankSim dataset, as shown in Fig.~\ref{SQNN_0}, the model achieves high accuracy under noise-free conditions \textcircled{\raisebox{-0.9pt}{1}}. However, as the noise level increases to 0.25 \textcircled{\raisebox{-0.9pt}{2}}, the performance drops significantly, particularly under depolarizing, phase damping, and amplitude damping noise. The model exhibits limited robustness, with minimal recovery across most noise types. Only the phase flip channel maintains some stability near \textcircled{\raisebox{-0.9pt}{3}}. Overall, accuracy remains within the 0.45–0.55 range even at moderate noise levels, suggesting the SQNN’s higher sensitivity to quantum perturbations and limited tolerance to decoherence effects.
Next, we analyze the VQC model on the European dataset. As shown in Fig.~\ref{VQC Noise}, the model achieves an accuracy of 0.90 under ideal, noise-free conditions \textcircled{\raisebox{-0.9pt}{1}}, demonstrating strong learning capability. As the noise level increases to 0.25 \textcircled{\raisebox{-0.9pt}{2}}, the accuracy gradually declines, especially under depolarizing and phase-damping noise, which disrupts quantum coherence and phase information. Amplitude damping noise results in more stable behavior, likely due to its directional projection toward $\ket{0}$, which may align with certain circuit designs. Interestingly, phase flip noise leads to a performance peak at \textcircled{\raisebox{-0.9pt}{3}}, possibly due to a regularization-like effect. At maximum noise levels \textcircled{\raisebox{-0.9pt}{4}}, accuracy across all noise types drops to the 0.4–0.55 range. Despite this decline, the VQC model demonstrates greater robustness compared to SQNN, benefiting from its optimized ansatz structure, expressive encoding, and effective classical post-processing.
This comparative analysis underscores the critical role of model architecture in determining quantum noise resilience. While the SQNN model is more susceptible to circuit-level perturbations, the VQC model maintains greater stability and adaptability, making it a more suitable candidate for deployment in real-world quantum applications such as fraud detection.
\subsection{Computational Complexity of QML Architectures}
From a theoretical perspective, the three QML architectures examined in this study, VQC, SQNN, and EQNN, share the same asymptotic computational complexity on quantum hardware. For a parametrized quantum circuit with $n$ qubits and depth $d$, a forward pass requires $O(d \cdot n)$ gates, while a gradient-based update with $p$ trainable parameters and $m$ measurement shots scales as $O(p \cdot m \cdot d \cdot n)$. Under these standard assumptions for variational quantum circuits, all three architectures belong to the same polynomial-time complexity class \cite{cai2022sample}.

Beyond this ideal setting, additional overhead arises from mapping logical circuits to the device’s native gate set and connectivity graph. Gate decomposition, routing, and SWAP insertion may increase circuit depth by a constant or topology-dependent factor, but these hardware effects do not modify the overall asymptotic scaling.

Classical simulation, in contrast, incurs exponential complexity: a state-vector simulator must track $2^n$ amplitudes \cite{zhou2020limits,miller2025phase2}, resulting in a cost of $O(2^n \cdot d)$ for all three architectures. Practical runtime differences therefore originate not from circuit depth or qubit count, but from how each architecture processes measurements. VQC and SQNN rely on the Sampler primitive to produce probability distributions, whereas EQNN uses the Estimator primitive to compute expectation values. Because each model applies the same feature map and ansatz, their gate-level complexity is identical; only constant-factor differences arise from measurement and output-mapping procedures such as parity mapping in VQC, multi-class probability construction in SQNN, or expectation-value evaluation in EQNN.

Since optimizer convergence depends on the landscape of the loss function, iteration counts are excluded from the asymptotic analysis. Overall, these considerations show that observed runtime differences across architectures arise mainly from measurement primitives, hardware-induced overhead, and classical post-processing rather than from intrinsic differences in computational complexity.
\subsection{Discussion}

The analytical results highlight the strong interaction between feature maps, ansatz configurations, and dataset characteristics in shaping model performance. A model’s ability to balance its performance metrics depends on how effectively the feature map encodes relevant relationships and how well the ansatz can exploit these encoded structures. These interactions vary across datasets and architectures, underscoring the need for configuration-aware model design.

Performance differences across architectures reflect their distinct output-processing mechanisms. VQC consistently benefits from expressive feature map–ansatz combinations, achieving high precision and stable F1-scores, particularly on the European dataset. SQNN demonstrates strong adaptability, excelling with configurations that balance computational simplicity and representational power, such as the Z feature map combined with Real Amplitudes on the BankSim dataset. In contrast, EQNN shows difficulty capturing complex patterns, likely due to structural limitations such as low circuit expressiveness, restricted feature mixing, and the constraints of mapping expectation values to class labels. Its sensitivity to configuration choices further indicates limited architectural flexibility compared to VQC and SQNN. Improving EQNN performance would require increasing circuit depth or expressivity, exploring alternative expectation-to-label mappings, or adopting task-specific loss functions that enhance sensitivity to minority-class patterns. These considerations highlight the importance of aligning architectural properties with appropriate encoding and variational strategies when deploying QML models for fraud detection.

The findings further indicate that entanglement-enhanced feature maps (e.g., ZZ, Pauli) can improve performance, but only when the model can effectively handle the added circuit complexity. SQNN, for instance, benefits substantially from such feature maps when paired with flexible ansatz like TwoLocal. Simpler feature maps, such as Z, frequently achieve strong results due to low depth, reduced noise sensitivity, and easier optimization, making them effective for less expressive models or noisy conditions.

Overall, the differences across feature maps and ansatz configurations align with theoretical principles related to circuit expressivity, entanglement structure, and the coupling between data encoding and variational layers. Configurations that enable richer expressibility and stronger feature mixing yield more stable and accurate models, while shallow or weakly entangling designs limit learning capacity. Although this study is empirical, the trends observed are consistent with established results on expressibility, trainability, and encoding efficiency.

These empirical observations are further supported by statistical evidence from the ANOVA analysis. Model type, feature map, and ansatz all exhibit statistically significant effects on F1-score, with model type producing the highest F-value. This confirms that the observed performance variability arises from meaningful architectural and configuration factors. In contrast, the dataset factor is not statistically significant, indicating that performance trends remain consistent across both the European and BankSim datasets.

To further contextualize these findings, the robustness trends under quantum noise offer additional insight into how architectural choices influence stability in realistic quantum environments. Models with well-structured variational layers demonstrate a stronger ability to preserve learned decision boundaries under perturbations, whereas architectures with limited expressiveness or weaker feature mixing degrade more rapidly. Across the simulated noise channels, depolarizing, phase damping, and amplitude damping exerted the strongest negative effects, while phase-flip noise was comparatively milder. Strategies such as reducing circuit depth, employing noise-adaptive ansatz designs, or integrating error-mitigation techniques may help improve robustness, although potentially at added computational cost.

This study demonstrates the importance of a strategic, data-driven approach to configuring QML models. The interplay between dataset properties, architectural structure, and feature map–ansatz combinations determines the achievable balance between computational efficiency and predictive accuracy. These insights can guide the design of more effective quantum and hybrid quantum models for financial fraud detection and similarly complex decision-making tasks.

\section{Conclusion\label{sec5}}
This study provides an insightful comparison of the performance of VQC, SQNN, and EQNN models in the context of financial fraud detection, highlighting their unique capabilities and limitations. By systematically exploring the impact of feature maps and ansatz configurations across two diverse datasets, the analysis highlights the critical factors that impact the effectiveness of QML models.
The results demonstrate that VQC and SQNN models are highly competitive in their ability to detect fraud, with VQC achieving the highest performance. In contrast, EQNN shows limited capacity to handle the complexity of the datasets, pointing to potential areas for architectural enhancement. The findings emphasize the significance of tailoring model configurations, including selecting feature maps and ansatz, to align with specific data characteristics and performance requirements.

Beyond comparative evaluation, this work makes two complementary contributions. First, a one-way ANOVA test verifies the statistical significance of the effects of model structure, dataset, feature map, and ansatz on performance. This adds methodological rigor and confirms that the observed differences are not due to random variation. Second, the robustness of the best-performing model for each dataset is evaluated under five types of quantum noise. The results show that they maintain acceptable performance levels under noisy conditions, supporting their potential deployment on NISQ devices.

Future research can build on these findings by exploring more expressive quantum architectures, deeper circuits, and improved feature encodings. The performance limitations observed in some models may result from factors such as limited circuit depth, insufficient ansatz expressiveness, or mismatches between configuration and data complexity. Further investigation is needed to isolate these factors and guide the development of more effective QML solutions for complex classification tasks.

\section*{Acknowledgment}
This work was supported in part by the NYUAD Center for Quantum and Topological Systems (CQTS), funded by Tamkeen under the NYUAD Research Institute grant CG008, and the Center for Cyber Security (CCS), funded by Tamkeen under the NYUAD Research Institute Award G1104.

\bibliographystyle{IEEEtran}

\bibliography{refs}

\end{document}